\documentclass[%
 reprint, 
nofootinbib, 
 amsmath,amssymb,
 aps,
 pra,
]{revtex4-2}
\usepackage{caption}
\usepackage{subcaption}
\usepackage{soul}
\usepackage[normalem]{ulem}
\usepackage{xcolor}
\usepackage{graphicx}
\usepackage{hyperref}
\hypersetup{
    colorlinks=true,
    linkcolor=blue,
    citecolor=magenta,
    filecolor=magenta,
    urlcolor=blue,
    pdftitle={}, 
    }
\allowdisplaybreaks
\newcommand{\be}{\begin{equation}}
\newcommand{\ee}{\end{equation}}
\newcommand{\bs}{\boldsymbol}
\newcommand{\mbf}{\mathbf}

\newcommand{\bx}{\mbf{x}}
\newcommand{\br}{\mbf{r}}

\newcommand{\bu}{\mbf{u}}

\newcommand{\ebold}{\mbf{E}}
\newcommand{\bbold}{\mbf{B}}
\newcommand{\ep}{\bs{\mathcal{E}}}
\newcommand{\bp}{\bs{\mathcal{B}}}

\newcommand{\z}{\hat{\mbf{z}}}

\newcommand{\bb}{\mbf{b}}
\newcommand{\bbot}{\mbf{b}_\bot}

\newcommand{\tbr}{\tilde{\br}}
\newcommand{\tr}{\tilde{r}}
\newcommand{\tphi}{\tilde{\varphi}}

\begin{document}

\preprint{APS/123-QED}

\title{Axion electrodynamics and the advection--diffusion structure of \\[3pt] TEM modes in hollow-conductor waveguides}

\author{Andr\'{e} H.\ Gomes}
\email{andre.h.gomes@ufes.br}
\affiliation{Departamento de F\'{i}sica, Universidade Federal do Esp\'{i}rito Santo, 29075-910,\\ Vit\'{o}ria, Esp\'{i}rito Santo, Brazil}%


\begin{abstract}
We present a systematic investigation of transverse electromagnetic (TEM) modes in hollow-conductor waveguides filled with a dielectric medium described by axion electrodynamics. While such modes do not propagate in conventional single-conductor waveguides, the coupling of the axion field to an external magnetic field enables their propagation. We show that the resulting TEM modes are governed by an advection--diffusion-type equation for the transverse electrostatic potential, establishing a direct mathematical connection between axion electrodynamics and advection--diffusion phenomena. We illustrate this connection through explicit solutions and field profiles for cylindrical waveguides, considering uniform and rigid rotational drift fields, both individually and in combination, as well as different values of the parameter controlling the advection-to-diffusion ratio (P\'{e}clet number). These results establish a transport-based framework for understanding how spatial variations of the axion-electrodynamics background shape the transverse electromagnetic fields in hollow-conductor waveguides.
\end{abstract}

\maketitle

\section{Introduction}

Axion electrodynamics extends Maxwell theory by coupling the electromagnetic field to an axion field $\theta$ \cite{Wilczek},
\be\label{action}
\mathcal{L}_\theta = -\frac{\kappa}{4} \theta(x)
F_{\mu\nu}\tilde F^{\mu\nu} = \kappa \theta(\bx,t) \ebold \cdot \bbold,
\ee
where $\kappa \equiv \varepsilon_0 c (\alpha/\pi) = e^2/4\pi^2\hbar$ in SI units, $\alpha$ is the vacuum fine-structure constant, and the metric signature is $(+,-,-,-)$. Although the pseudoscalar field $\theta$ underlying this construction is motivated by the QCD axion arising from the Peccei--Quinn mechanism for the strong $CP$ problem in particle physics \cite{weiberg-axion,wilczek-axion,RevModPhys.axions}, the same $\theta F \tilde{F}$ structure also occurs in effective descriptions of condensed-matter systems \cite{axion-cond-mat}. The modified Maxwell equations contain axion-induced charge and current densities that, depending on the physical realization, can be associated with anomalous Hall and chiral responses in material media \cite{Nagaosa-RMP-2010,kharzeev,chiralmatter}.

For a prescribed, nondynamical axion field, the theory can also be formulated in terms of the background four-vector
\be\label{background}
b^\mu(x) \equiv \partial^\mu\theta(x),
\ee
with the corresponding Lagrangian\footnote{Lagrangians \eqref{action} and \eqref{CFJ} differ by the total four-divergence $\frac{\kappa}{2} \partial_\mu (\theta A_\nu \tilde{F}^{\mu\nu})$ and therefore yield identical bulk equations of motion. For the perfectly conducting boundaries considered here, the spatial surface contribution vanishes by the boundary conditions $\mbf{n} \times \ebold=0$ and $\mbf{n} \cdot \bbold = 0$ (cf. Sec.~\ref{sec:TEM}). For more general spatial boundary conditions, this surface contribution need not vanish; see, e.g., Refs.~\cite{BC-2011,BC-2015,BC-2016,BC-2018}}
\be\label{CFJ}
\mathcal{L}_\theta= \frac{\kappa}{2} b_\mu
A_\nu \tilde{F}^{\mu\nu}
= \frac{\kappa}{2}[ b_0 \mbf{A}\cdot\bbold + \bb \cdot (\mbf{A} \times \ebold - \phi \bbold)].
\ee
Despite its explicit dependence on the gauge potential $A^\mu=(\phi/c,\mbf{A})$, the Lagrangian is gauge invariant, and electric charge is therefore conserved.\footnote{Under a gauge transformation, $\mathcal{L}_\theta$ changes by a term proportional to $\tilde{F}^{\mu\nu} (\partial_\mu b_\nu - \partial_\nu b_\mu)$, which vanishes identically by \eqref{background}.} This theory was introduced in the context of Lorentz-violating electrodynamics for fixed $b^\mu$, when the axion derivative $\partial^\mu\theta$ is constant, and is commonly referred to as Maxwell-Carroll-Field-Jackiw electrodynamics \cite{cfj,sme01,sme02,jackiw}. The vector $b^\mu$ selects a preferred spacetime direction, thereby defining an anisotropy that breaks Lorentz symmetry. This framework has also been applied to the effective description of condensed-matter systems \cite{LSV-Semimetal,urrutia-weyl,Large-LV-WS,weyl-superconductor,rogeres}, and particularly in topological materials \cite{Zhang-Nature-Physics-2010,Sekine-JAP-2021,RevModPhys-2018,Deng-PRB104-075202-2021}, including Weyl semimetals, where the temporal and spatial components of the medium anisotropy $b^\mu$ are related to the energy and momentum-space separations of the Weyl nodes, respectively \cite{Burkov-PRB85-165110-2012,RevModPhys-2018,elight}.

Axion electrodynamics has been investigated in a variety of optical and electromagnetic settings. At interfaces between media with different axion couplings, it modifies the Fresnel reflection coefficients and the conditions for Brewster angles, and gives rise to distinctive Goos--Hänchen shifts and polarization rotation \cite{2009-optical-chang,2014-optical-huerta}. Optical proposals have also been developed to measure the quantized topological magnetoelectric response of three-dimensional topological insulators \cite{2010-optical-TME}. In Weyl semimetals, axion terms associated with broken inversion and time-reversal symmetries affect electromagnetic-wave propagation, transmission, and Faraday rotation, leading to nonreciprocity, dichroism, and birefringence \cite{2023-optical-filtration-reflection}. Related studies of chiral media with anomalous Hall response have reported circular birefringence, rotatory power with double sign reversal, unusual Kerr rotation and ellipticity, and anomalous reflectance exceeding unity \cite{ufma-rotatory,2025-anomalous-reflectance}. Beyond optical propagation, Green-function methods have been developed to characterize static electromagnetic fields and boundary effects at planar, spherical, and cylindrical interfaces between regions with different axion couplings \cite{BC-2015,BC-2016,BC-2018}.

\begin{figure}[htb]
	\centering
	\includegraphics[width=.95\linewidth]{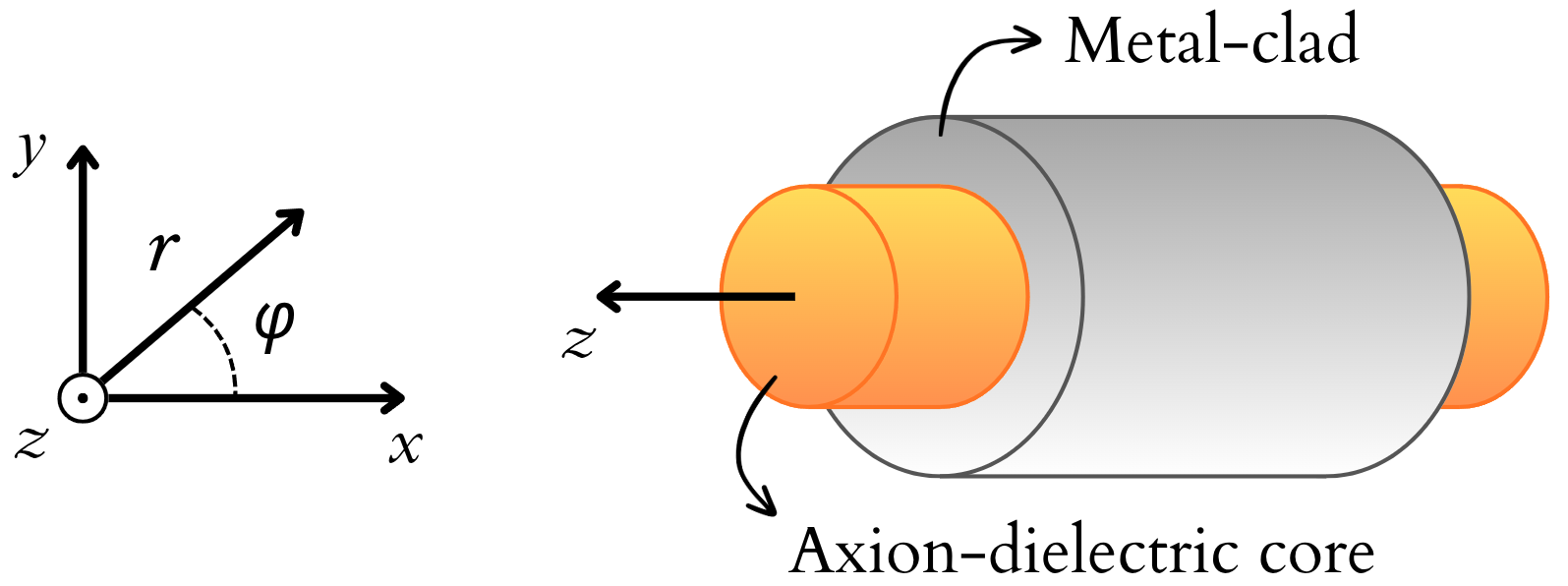}
	\caption{Cylindrical metal-clad waveguide (gray) with an axion-dielectric core (orange).}
	\label{fig:waveguide}
\end{figure}

Motivated by the use of axion electrodynamics to describe the low-energy electromagnetic response of Weyl semimetals, we recently showed that a hollow-conductor waveguide filled with an axion medium (Fig.~\ref{fig:waveguide}) can support transverse electromagnetic (TEM) modes that are absent in conventional single-conductor waveguides \cite{andre-winder}. This mechanism relies on two conditions: (i) an appropriate choice of the axion background, such that the transverse contributions of the chiral magnetic and Hall-like currents cancel, allowing a self-consistent TEM configuration; and (ii) an externally applied static, uniform magnetic field, which activates the medium-induced source terms required to sustain nontrivial transverse fields.

The possibility of modifying conventional waveguide propagation through axion and topological magnetoelectric responses has also been investigated in other settings. In particular, exact TEM solutions have been obtained in topological-insulator geometries involving spatial variations of the topological magnetoelectric parameter, with potential applications to optical fibers \cite{2024-optical-TEM}. Related studies have examined magnetoelectric effects in topological-insulator waveguides and guided electromagnetic modes in Weyl-semimetal structures \cite{Crosse2017,Filipini2025,Halterman2019,Peluso2025}. These works establish a broader context in which topological and axion-induced electromagnetic responses modify the propagation and polarization properties of guided modes. The present work addresses a complementary aspect of this problem, focusing on the transverse structure of axion-enabled TEM fields and, in particular, on the role of spatial variations of the axion background.

In our previous work \cite{andre-winder}, we restricted the analysis to configurations in which the axion medium effectively acts as a source. Here, we relax this restriction and consider more general transverse variations of the axion background $\theta$. In particular, the associated transverse anisotropy $\bbot$ defines an effective drift field $\bu=\bbot\times\z$. We show that the resulting TEM modes are governed by a steady advection--diffusion-type equation for the transverse electrostatic potential. This formulation provides a direct mathematical connection between axion electrodynamics and transport phenomena, in which the spatial structure of the axion background determines the effective drift field and thereby controls the transverse electromagnetic profiles.

We investigate this correspondence through analytical and asymptotic solutions for representative configurations, including pure diffusion, uniform drift, rigid rotational drift, and the combined effects of uniform and rotational drifts. We consider circular waveguides, as well as an elliptical cross section to investigate the interplay between rotational advection and geometrical asymmetry. The solutions are analyzed in both diffusion-dominated and advection-dominated regimes, characterized by the P\'{e}clet number, and reveal the development of anisotropic field profiles, boundary-layer accumulation, and distinct bulk and boundary structures. We also identify the conditions under which nonuniform irrotational drift fields admit a reduction to a modified Helmholtz equation. Together, these results provide a transport-based framework for characterizing the transverse structure of axion-enabled TEM modes and clarify how the spatial configuration of the axion background shapes their electromagnetic fields.

The remainder of the paper is organized as follows. Section~\ref{sec:TEM} introduces the physical setting of TEM propagation in hollow-conductor waveguides and establishes the conditions for its occurrence. Section~\ref{sec:governing-equation} then derives the governing equation for the electrostatic potential and maps it onto an advection--diffusion problem, addressing its conservation form, incompressible drift fields, a P\'{e}clet-like number, and the role of the external magnetic field. Section~\ref{sec:solutions} applies this framework to cylindrical waveguides, considering pure diffusion, uniform drift, rigid rotational drift, and combined uniform and rigid rotational drifts, as well as the solutions in the low- and high-P\'{e}clet-number regimes. At last, in Sec.~\ref{sec:conclusion} we present our concluding remarks.

\section{TEM propagation in hollow-conductor waveguides in axion electrodynamics}
\label{sec:TEM}

This section reviews the theoretical framework developed in Ref.~\cite{andre-winder} for TEM propagation in hollow-conductor waveguides filled with an axion medium with uniform, frequency-independent permittivity $\varepsilon$ and permeability $\mu$. For convenience, we adopt the complex representation for both the electromagnetic and axion fields, with the physical fields given by the real parts of their respective complex representations.

The axion background modifies electromagnetic propagation through Maxwell's equations supplemented by the axion-induced terms derived from \eqref{action} or, equivalently, from \eqref{CFJ}. In the absence of externally imposed free charge and current densities, these equations read
\begin{align}
&\nabla \cdot \ebold = \frac{\kappa}{\varepsilon} \bb \cdot \bbold,
\displaybreak[1]\\[5pt]
& \nabla \times \ebold + \frac{\partial \bbold}{\partial t} = 0,
\displaybreak[1]\\[5pt]
& \nabla \cdot \bbold = 0,
\displaybreak[1]\\[5pt]
& \nabla \times \bbold - \frac{1}{v^2} \frac{\partial\ebold}{\partial t} = \mu \kappa (b_0 \bbold - \bb \times \ebold),
\end{align}
where $b_0 = \partial_t\theta$ and $\bb=-\nabla\theta$ can be spacetime dependent. In the inhomogeneous equations, the axion terms act as material-induced sources; namely,
\be
\rho_\text{Hall} \equiv \kappa \bb \cdot \bbold,
\qquad
\mbf{J}_\text{Hall} \equiv - \kappa \bb \times \ebold
\ee
defines Hall-like charge and current densities, respectively, and
\be
\mbf{J}_\text{chiral} \equiv \kappa b_0 \bbold
\ee
defines a chiral current density. The homogeneous Maxwell equations remain unaltered, since the axion term does not introduce magnetic monopoles or otherwise modify the kinematic structure of the electromagnetic field \cite{ney-ufv}.

In the source-free setup, TEM modes are described by the following time-harmonic ansatz for a transverse wave propagating along the guide axis:
\begin{align}\label{ansatz}
& \ebold_\text{TEM}(\bx,t) = \ep(x,y) e^{i(kz-\omega t)}, \qquad \ep\cdot\z = 0,
\\[5pt]
& \bbold_\text{TEM}(\bx,t) = \bp(x,y) e^{i(kz-\omega t)}, \qquad \bp\cdot\z = 0.
\end{align}
The guide axis is taken along the $\z$-direction, and the transverse radial vector is denoted by $\br$, where $\br=(x,y) = (r\cos\varphi, r\sin\varphi)$ in Cartesian and cylindrical coordinates, respectively. The transverse fields $\ep$ and $\bp$ are assumed to satisfy the homogeneous boundary conditions
\be\label{BCs}
\hat{\mbf{n}} \times \ep|_\mathcal{S} = 0,
\qquad
\hat{\mbf{n}} \cdot \bp|_\mathcal{S} = 0,
\ee
where $\hat{\mbf{n}}$ is the unity normal vector at the guide walls $\mathcal{S}$, assumed to be a perfect conductor.

The TEM ansatz is compatible with the modified Maxwell equations provided that the following relations hold \cite{andre-winder}:
\be\label{conditions-for-TEM}
\bp = \frac{\z \times \ep}{v},
\qquad
\omega = v k,
\qquad
b_0 = v b_z,
\ee
where $v=1/\sqrt{\mu\varepsilon}$ is the effective speed of light in the material. The first two relations are standard for TEM propagation and imply nondispersive propagation at speed $v$ within the assumed frequency-independent constitutive model. The third relation is specific to axion electrodynamics and requires a suitably tuned axion medium; it follows from the transverse projections of Faraday’s and Ampère--Maxwell’s equations. Physically, it enforces the cancellation
\be
\mbf{J}_\text{Hall}^\bot + \mbf{J}_\text{chiral}^\bot = 0,
\ee
where
\be
\mbf{J}_\text{Hall}^\bot = - \kappa b_z \z \times \ep,
\qquad
\mbf{J}_\text{chiral}^\bot = \kappa b_0 \bp
\ee
are the transverse Hall-like and chiral current densities. If these currents were not balanced, they would generate longitudinal fields and thus destroy the TEM character.

Using the relations in \eqref{conditions-for-TEM}, Maxwell's equations reduce to the following pair of equations for the transverse electric field:
\be
\nabla \cdot \ep - \frac{\kappa}{v\varepsilon}(\bbot \times \z) \cdot \ep = 0,
\qquad
\nabla \times \ep = 0,
\ee
where $\bbot$ is the medium anisotropy in the transverse plane. Equivalently, since $\bp = \z \times \ep/v$, they can be expressed as a pair of equations for the transverse magnetic field:
\be
\nabla\times \bp - v\mu\kappa(\bbot \cdot \bp) \z = 0,
\qquad
\nabla \cdot \bp = 0.
\ee
This system defines a two-dimensional problem: the differential operators act on the transverse fields $\ep(x,y)$ and $\bp(x,y)$, while $\bbot$ lies in the $xy$-plane, perpendicular to the waveguide axis. The terms involving $\bbot$ describe Hall-like charge and axial current densities induced by the coupling between the axion medium and the TEM-wave components $\ep$ and $\bp$,\footnote{Namely, $\rho_\text{Hall} = \kappa(\bbot \times \z) \cdot \ep/v$ and $\mbf{J}_\text{Hall}^\parallel = (v\kappa \bbot \cdot \bp) \z$.} respectively. These coupling terms determine the transverse field profiles and can lead to field accumulation under suitable boundary conditions, as discussed in following sections. However, TEM propagation is not possible when the axion medium is tightly enclosed by a single conducting waveguide, as in Fig.~\ref{fig:waveguide}. The conducting boundary imposes a constant electric potential, while the differential equation for the potential forbids local maxima or minima in the interior. The potential must therefore vanish identically, as must the transverse fields (cf. Sec.~\ref{sec:external-field}). This restriction is analogous to that encountered in waveguides filled with conventional media \cite{Griffiths2013,Jackson1999}.

This restriction on TEM propagation can be circumvented in a hollow-conductor waveguide by applying a static, uniform external magnetic field,
\be
\bbold_\text{ext} = B_\text{ext}\z,
\ee
throughout the bulk material filling the guide, provided that the prescribed axion field is of the form\footnote{In contrast, Ref.~\cite{andre-winder} considers $\theta = b_0 t - \bb\cdot\bx$ in our conventions, which corresponds to the small-phase limit, $|kz-\omega t| \ll 1$, of Eq.~\eqref{axion} when $\theta_{xy}=-\bbot\cdot\br$, provided that $b_0=\omega\delta\theta$ and $b_z=k\delta\theta$. The equivalence holds up to an additive constant, which is unobservable in the present bulk context.}
\be\label{axion}
\theta(\bx,t) = \theta_{xy}(x,y) + i \delta\theta e^{i(kz-\omega t)},
\ee
where $\delta\theta$ is the real amplitude of the axion perturbation propagating along the waveguide axis. The temporal and axial components of the medium anisotropy $b^\mu$ satisfy
\be\label{b0bz}
b_0 = \omega \delta\theta e^{i(kz-\omega t)} = v b_z,
\ee
as required. The position-dependent medium anisotropy in the transverse plane is identified as
\be
\bbot \equiv -\nabla\theta_{xy},
\ee
where $\theta_{xy}$ contains all the axion-background dependence on coordinates in the $xy$-plane.

In this case, Maxwell's equations for the fields
\be
\ebold =\ebold_\text{TEM},
\qquad
\bbold = \bbold_\text{TEM} + B_\text{ext}\z
\ee
become\footnote{In Ref.~\cite{andre-winder}, this set of equations is obtained by accounting for free charges and the material polarization and magnetization, under the assumption that light propagates within the optically transparent frequency window of the bulk material and spatial variations in the material's magnetization are negligible.}
\be\label{electric-PDE}
\nabla \cdot \ep - \frac{\kappa}{v\varepsilon}(\bbot \times \z) \cdot \ep = \frac{\kappa}{\varepsilon}|b_z| B_\text{ext},
\qquad
\nabla \times \ep = 0,
\ee
or, equivalently,
\be\label{magnetic-PDE}
\nabla \times \bp - v\mu\kappa(\bbot \cdot \bp)\z = \mu \kappa |b_0| B_\text{ext}\z,
\qquad
\nabla \cdot \bp = 0.
\ee
Unlike $\bbot$, which couples only to the TEM fields, $b_0$ and $b_z$ couple to the externally applied magnetic field, giving rise to a Hall-like charge density and a longitudinal chiral current density.%
\footnote{Explicitly, $\rho_\text{Hall}^\text{(ext.)} = \kappa b_z B_\text{ext}$ and  $\mbf{J}_\text{chiral}^{\parallel\text{(ext.)}} = (\kappa b_0 B_\text{ext})\z$. These terms inherit their spacetime dependence from $b_0$ and $b_z$, as defined in \eqref{b0bz}.}
The former acts as a source for the transverse electric field $\ep$, whereas the latter sources the transverse magnetic field $\bp$. The common phase factor $e^{i(kz-\omega t)}$ of these induced sources, together with the relation $b_0=(\omega/k)b_z$, ensures that they form a charge-conserving distribution propagating along the waveguide axis. When phase-matched to the TEM dispersion relation, these sources enable a propagating TEM solution.

At this point, it is interesting to note the many consequences of the conditions in \eqref{conditions-for-TEM}. They require the transverse Hall-like and chiral currents, $\mbf{J}_\text{Hall}^\bot$ and $\mbf{J}_\text{chiral}^\bot$, to cancel, thereby eliminating the longitudinal wave components. They also connect the Hall-like charge density induced by the TEM wave to its axial current through $\mbf{J}_\text{Hall}^\parallel = v \rho_\text{Hall} \z$. Similarly, the externally induced Hall-like charge and longitudinal chiral current densities are related by $\mbf{J}_\text{chiral}^{\parallel\text{(ext.)}} = v \rho_\text{Hall}^\text{(ext.)} \z$. Combined with charge conservation, this relation implies that the induced Hall-like charge density propagates along $\z$ at the same speed $v$ as the TEM mode.\footnote{For axion electrodynamics applied to the optical response of Weyl semimetals, this establishes a correspondence between electromagnetic propagation and the charge dynamics induced by the Weyl-semimetal response. A related microscopic connection between electromagnetic propagation and Weyl-fermion dynamics was found in Ref.~\cite{PRB-Unconventional-Mode}, where the unconventional electromagnetic mode propagates at the Weyl-fermion Fermi velocity $v_F$ in the low-energy regime. Here, the corresponding velocity is instead the macroscopic electromagnetic velocity $v=1/\sqrt{\mu\varepsilon}$.}

Subject to the boundary conditions \eqref{BCs}, equations \eqref{electric-PDE} and \eqref{magnetic-PDE} determine the transverse fields and, together with the ansatz \eqref{ansatz}, fully specify the TEM modes propagating in a hollow-conductor waveguide. In Ref.~\cite{andre-winder}, we established nontrivial solutions for multiple waveguide configurations under the restriction that the charge and current densities induced by the coupling between the axion medium and the TEM wave vanish, $\bbot \cdot \bp \propto (\bbot \times \z) \cdot \ep = 0$ in Eqs.~\eqref{electric-PDE} and \eqref{magnetic-PDE}. Under this restriction, the axion medium acts only as a source for the TEM modes through the external field. As shown below, relaxing this condition and allowing the axion-induced Hall-like couplings to contribute to the transverse field structure establishes a direct correspondence between TEM propagation in a hollow-conductor waveguide and a driven steady-state advection--diffusion problem.

\section{Governing equation and advection--diffusion mapping}
\label{sec:governing-equation}

Using $\bp = \z \times \ep/v$, Eqs.~\eqref{electric-PDE} and \eqref{magnetic-PDE} for the transverse fields are equivalent. Introducing $\ep=-\nabla\phi(x,y)$ reduces them to a single equation for the electrostatic potential in the transverse plane,
\be
\nabla^2\phi - \frac{\kappa}{v\varepsilon}(\bbot \times \z) \cdot \nabla \phi = - \frac{\kappa}{v\varepsilon}|b_0| B_{\text{ext}}.
\ee
At this point, it is convenient to absorb the factor $\kappa/v\varepsilon$ into the definition of the axion field, so that
\be
\frac{\kappa}{v\varepsilon} b^\mu \to b^\mu.
\ee
After this rescaling, $b^\mu$ has dimensions $[b_0]=[v][L]^{-1}$ and $[\bb]=[L]^{-1}$. The governing equation for $\phi$ assumes the simpler form
\be\label{adv-diff-b}
\nabla^2\phi - (\bbot \times \z) \cdot \nabla \phi = - |b_0| B_{\text{ext}}.
\ee
Equation \eqref{adv-diff-b} can be identified with the steady-state advection--diffusion equation for a transported scalar field,
\be\label{advection-diffusion}
D\nabla^2\phi-\bu\cdot\nabla\phi=-S_0,
\ee
under the correspondence
\be
D=1,
\qquad
\bu = \bbot \times \z,
\qquad
S_0 = |b_0| B_{\text{ext}}.
\ee
We use the term \textit{advection-diffusion equation} to describe the mathematical structure of the governing equation \eqref{adv-diff-b}, following the standard terminology of continuum transport theory; see, e.g., Ref.~\cite{LeVeque2004}. The vector field $\bu$ is referred to as the \textit{drift field}: it plays the role of a transport velocity in the advection-diffusion analogy, but does not represent a physical fluid velocity in this system. In this analogy, $\bbot \times \z$ acts as the drift field associated with Hall-like transport in the transverse plane, whereas $|b_0| B_\text{ext}$ plays the role of the uniform source term $S_0$, associated with the externally induced Hall-like charge density. The diffusion-like coefficient $D$ is dimensionless and normalized to unity.

The TEM fields are thus governed by a driven steady-state advection--diffusion equation subject to the Dirichlet boundary condition
\be\label{dirichlet}
\phi(x,y)|_\mathcal{C} = 0,
\ee
Here, $\mathcal{C}$ denotes the boundary of the waveguide cross section. This condition follows from the perfect-conductor boundary conditions \eqref{BCs} and closes the transverse problem, yielding a unique steady-state field profile.

\subsection{Conservation form}
\label{sec:conservation-form}

Equation \eqref{adv-diff-b} can be equivalently written as a stationary continuity-like equation,
\be\label{conservation-form}
\frac{\partial\phi}{\partial t} = \nabla \cdot \mbf{J}_{\phi} - S_0 = 0,
\ee
with
\be
\mbf{J}_{\phi} = - \nabla \phi + \bu \phi.
\ee
The first term represents the diffusive flux of the potential, while the second describes its advective, Hall-like transport by the drift field $\bu$. This representation is commonly known as the conservation form of the advection--diffusion equation \cite{Wesseling2001,LeVeque2004}. Equivalence with Eq.~\eqref{adv-diff-b} requires $\bu$ to be solenoidal,
$\nabla\cdot\bu=0$, which is automatically satisfied by the axion backgrounds $\theta_{xy}$ considered here. In fact, using
\be\label{drift-field}
\bu=\bbot\times\z = - \nabla \times (\theta_{xy} \z),
\ee
we obtain
\be
\nabla\cdot\bu=0
\ee
for sufficiently regular $\theta_{xy}$.

The drift field, however, need not be irrotational. For the above definition of $\bu$,
\be\label{vorticity}
\nabla\times\bu
=-\nabla^2\theta_{xy} \z.
\ee
The drift is therefore irrotational only if $\theta_{xy}$ is harmonic, $\nabla^2\theta_{xy}=0$. Thus, $\theta_{xy}$ generates an incompressible drift field whenever it is sufficiently regular; the additional requirement of vanishing vorticity imposes the condition that $\theta_{xy}$ be harmonic.

\subsection{P\'{e}clet-like number}

The competition between advection and diffusion in a steady advection--diffusion problem is characterized by the P\'{e}clet number \cite{Wesseling2001,Leal2007},
\be
\text{Pe}=\frac{uL}{D},
\ee
where $u$ is a characteristic value of drift speed, $D$ is the diffusion coefficient, and $L$ is the characteristic length scale of the domain. In the present dimensionless formulation, $D=1$ and $u=|\bbot\times\z|=b_\bot$ is the magnitude of the transverse drift field or, more generally, its characteristic value when the magnitude of $\bbot$ varies across the transverse plane, so that
\be
\text{Pe}=b_\bot L.
\ee
This dimensionless parameter compares the characteristic advective transport rate, $u/L$, with the diffusive rate, $D/L^2$. Accordingly, $\text{Pe}\ll1$ defines a diffusion-dominated regime, in which the Laplacian term controls the transverse potential profile. Conversely, for $\text{Pe}\gg1$, the system is advection-dominated and transport by the Hall-like drift becomes increasingly important. The crossover between these regimes occurs for $\text{Pe}\sim1$.

The same interpretation follows from the relative scaling of the two terms in Eq.~\eqref{adv-diff-b}. If $\phi$ varies over a characteristic transverse length $L$, then
\be
\nabla^2 \phi \sim \frac{\phi}{L^2},
\qquad
(\bu \cdot \nabla)\phi \sim \frac{u\phi}{L},
\ee
and hence
\be
\frac{|(\bu \cdot \nabla)\phi|}
{|\nabla^2 \phi|} \sim uL = \text{Pe}.
\ee
Hence, $\text{Pe}$ directly measures the relative magnitude of the Hall-like drift and diffusion terms in the governing equation.

For the geometries considered here, $L$ is determined by the transverse dimensions of the waveguide. Increasing either the characteristic magnitude of the transverse medium anisotropy or the waveguide cross-sectional size therefore increases $\text{Pe}$, driving the system toward the advection-dominated regime. For sufficiently small $b_\bot L$, by contrast, Hall-like transport provides only a perturbative correction to the transverse potential profile, and the solution approaches the purely diffusive limit.

\subsection{Necessity of an external magnetic field}
\label{sec:external-field}

The source term in Eq.~\eqref{adv-diff-b} plays a qualitatively distinct role in determining the transverse field. In particular, under the homogeneous Dirichlet boundary condition \eqref{dirichlet} imposed by the perfectly conducting waveguide walls, the advection--diffusion equation admits only trivial solution in the absence of sources in a simply connected region. Consequently, a nonzero source is required to sustain a nonvanishing electrostatic potential in a hollow-conductor waveguide.

To establish this result, let $A$ denote the waveguide cross section, bounded by $\mathcal{C}$, and consider
\be
\nabla^2 \phi - \bu \cdot \nabla \phi = -S_0,
\qquad
 \left.\phi \right|_\mathcal{C} = 0,
\ee
where $\nabla\cdot\bu=0$. First, consider the source-free case ($S_0=0$) in a simply connected region $A$. Multiplying the advection-diffusion equation \eqref{advection-diffusion} by $\phi$ and integrating over $A$ gives
\be
\int_{A} \phi \nabla^2 \phi \,dA
- \int_{A} \phi \bu \cdot \nabla \phi \,dA = 0.
\ee
Integration by parts of the Laplacian yields
\be
\int_{A} \phi \nabla^2 \phi \,dA
= \oint_\mathcal{C} \phi (\nabla\phi)\cdot \hat{\mbf{n}}d\ell - \int_{A} |\nabla\phi|^2 \,dA
\ee
while the advective contribution can be written as
\begin{align}
\int_{A}\phi \bu \cdot \nabla \phi \,dA
& = \frac{1}{2} \int_{A} \bu \cdot \nabla(\phi^2) \,dA
\nonumber\\[5pt]
& = \frac{1}{2} \oint_\mathcal{C} \phi^2 \bu \cdot \hat{\mbf{n}} d\ell
- \frac{1}{2} \int_{A}(\nabla\cdot\bu) \phi^2 \,dA.
\end{align}
Both boundary terms vanish because $\phi=0$ on $\mathcal C$. The interior contribution involving $\nabla\cdot\bu$ vanishes because $\bu$ is a solenoidal vector field (Sec.~\ref{sec:conservation-form}). Therefore,
\be
\int_{A} |\nabla\phi|^2 \,dA = 0,
\ee
which implies $\nabla\phi=0$ throughout the cross-sectional area $A$. Since this region is simply connected, $\phi$ must then be constant, and the Dirichlet condition fixes this constant to zero. Thus, the problem with $S_0=0$ possesses only the trivial solution $\phi=0$. In particular, the Hall-like drift cannot by itself generate a nonzero transverse potential when the source term vanishes.

The same conclusion can be expressed through the integrated conservation law. From \eqref{conservation-form}, there follows
\be
\oint_\mathcal{C} \mbf{J}_{\phi} \cdot \hat{\mbf{n}} d\ell
= S_0 A,
\qquad
\mbf{J}_\phi = -\nabla\phi + \phi\bu.
\ee
Since $\phi=0$ on $\mathcal{C}$, the advective contribution to the boundary flux vanishes, resulting in
\be
\oint_\mathcal{C} (\nabla\phi)\cdot \hat{\mbf{n}}d\ell = - S_0 A.
\ee
The source therefore establishes a nonzero net boundary flux of the diffusive part of $\mbf{J}_{\phi}$. Since $\ep=-\nabla\phi$, this relation is equavalent to Gauss' law for the two dimensional problem in the transverse plane. When $S_0=0$, this driving mechanism is absent, and the homogeneous Dirichlet condition forces the entire solution to vanish.

In the present electromagnetic problem considered here,
\be
S_0 = |b_0| B_{\text{ext}},
\ee
so that, provided $|b_0|\neq0$, a nonvanishing TEM solution requires
\be
B_{\text{ext}}\neq0.
\ee
The external magnetic field therefore not only controls the magnitude of the TEM fields but also provides the source required to sustain a nontrivial transverse electrostatic potential under the perfectly conducting boundary condition. Thus, a nonzero external field is necessary for the nontrivial steady-state TEM configuration described by Eq.~\eqref{adv-diff-b}.

\section{Solutions}
\label{sec:solutions}

In this section, we obtain the electrostatic potential for representative advection--diffusion problems. We consider four cases: pure diffusion, uniform drift, rigid rotational drift, and combined uniform and rigid rotational drift. In all cases, the waveguide cross section is assumed to be cylindrical and rotationally symmetric, except in the case of rigid rotation, for which an elliptical cross section is also considered. Although analytical expressions for the electrostatic potential are derived, the corresponding transverse fields are omitted for brevity and are instead represented in the field plots.

\subsection{Pure diffusion}
\label{sec:pure-diffusion}

In the absence of a drift field ($\bu=0$), or when the potential is constant along the drift lines ($\bu \cdot \nabla\phi=0$), the electrostatic potential is governed by pure diffusion,
\be
\nabla^2 \phi = - S_0.
\ee
For a cylindrical waveguide of radius $R$ subject to a Dirichlet boundary condition, the solution is
\be\label{diffusion-disc-solution}
\phi(r) = \frac{S_0}{4}(R^2-r^2).
\ee
The corresponding transverse electric and magnetic fields are
\begin{align}
\ep(\br) = \frac{1}{2} S_0 r \hat{\br},
\qquad
\bp(\br) = \frac{1}{2v} S_0 r \hat{\bs{\varphi}}.
\end{align}
These solutions are rotationally symmetric, as expected from the symmetry of the diffusion equation and the boundary conditions. The corresponding field patterns are shown in Fig.~\ref{fig:uniform-DIFF}, for which the P\'{e}clet number is $\text{Pe}=0$. Solutions for other cross-sectional geometries, particularly rectangular and slab waveguides, are given in Ref.~\cite{andre-winder}.

\subsection{Uniform drift}
\label{sec:uniform-drift}

Advection generally breaks rotational symmetry by selecting a preferred spatial direction and displacing the diffusive profile. To illustrate this effect, we consider the same setting as above, now with a uniform drift field
\be
\bu=u\hat{\bu},
\qquad \text{(fixed vector)}
\ee
where $u$ is constant and $\hat{\bu}$ is a unit vector pointing along a fixed preferred direction.  Recalling that $\bu = \bbot \times \z$, this drift corresponds to the uniform transverse anisotropy
\be
\bbot = \z \times \bu.
\ee
Since $\bbot = - \nabla \theta_{xy}$, where $\theta_{xy}$ is the in-plane axion background, the background configuration that produces such a drift is
\be\label{axion-uniform-drift}
\theta_{xy}^{(u)} = - (\z \times \bu) \cdot \br = xu_y - yu_x.
\ee
For instance, for a drift field along the $x$-axis, this axion field becomes $-uy$, corresponding to a transverse anisotropy $u\hat{\mbf{y}}$. In the following, we set $\bu=u\hat{\mbf{x}}$ for simplicity.

For uniform drift, a particular solution $\phi_p$ of the advection--diffusion equation \eqref{advection-diffusion} is readily obtained:
\be
\phi_p(\br) = \frac{S_0}{u^2} \bu\cdot\br = \frac{S_0}{u}r\cos\varphi.
\ee
Writing the full solution as $\phi = \phi_h + \phi_p$ and applying the standard exponential transformation to the homogeneous part,
\be
\phi_h(\br) = \psi(\br)e^{\bu\cdot\br/2}
\ee
reduces the problem to solving the modified Helmholtz equation in the corresponding geometry,
\be\label{helmholtz}
\nabla^2\psi - \tfrac{1}{4}u^2 \psi = 0.
\ee
For a uniform drift field, the exponential factor in $\phi$ produces an asymmetric localization of the stationary solution, analogous to boundary accumulation in advection--diffusion problems. Since $\phi$ satisfies a Dirichlet boundary condition at the waveguide wall $\mathcal{C}$, the corresponding condition for $\phi_h$ is $\phi_h|_\mathcal{C}=-\phi_p|_\mathcal{C}$. Thus, the boundary condition for $\psi_h$ is
\be
\psi_h|_\mathcal{C} =  \left. \left[-\frac{S_0}{u^2}\bu\cdot\br e^{-\bu\cdot\br/2} \right] \right|_\mathcal{C}.
\ee
For the circular geometry considered here, this condition is imposed at $r=R$.

Since the domain is a circle centered at the origin and the boundary condition is an even function of $\varphi$, the sine terms vanish from the general solution of the modified Helmholtz equation that remains finite at the origin:
\be\label{u-drift-general-solution-helmholtz}
\psi_h(r, \varphi) = \sum_{n=0}^\infty  A_n I_n \left(\tfrac{1}{2}ur \right) \cos(n\varphi),
\ee
where $I_n$ denotes the modified Bessel function of the first kind of order $n$ \cite{arfken}.

The coefficients $A_n$ are obtained by matching this solution to the boundary condition
\be\label{u-drift-BC-psi}
\psi_h(R, \varphi) = -\frac{S_0R\cos\varphi}{u} e^{-\tfrac{1}{2}u R \cos\varphi}.
\ee
To this end, we use the Jacobi-Anger expansion,
\begin{align}
e^{iz\cos\varphi}
& = \sum_{n=-\infty}^\infty i^n J_n(z)e^{in\varphi}
\nonumber\\[5pt]
& = J_0(z) + 2 \sum_{n=1}^\infty i^n J_n(z)\cos(n\varphi),
\end{align}
where $J_n$ is the Bessel function of the first kind, and the second equality follows from the symmetry relation $J_{-n}(z)=(-1)^n J_n(z)$. Substituting $z\to iz$ and using $J_n(iz)=i^n I_n(z)$, we obtain
\be
e^{-z \cos\varphi} = I_0(z) + 2 \sum_{n=1}^\infty  (-1)^n I_n(z) \cos(n\varphi).
\ee
Differentiating both sides with respect to $z$ and using the identity $I_0'(z) = I_1(z)$ yields an expression for $\cos\varphi e^{-z \cos\varphi}$. Evaluating this expression at $z=uR/2$ allows us to express \eqref{u-drift-BC-psi} as a Fourier series in $\cos(n\varphi)$. Matching the resulting series to \eqref{u-drift-general-solution-helmholtz} at $r=R$ gives
\begin{align}
& A_0 = \frac{S_0 R}{u} \frac{I_1(\tfrac{1}{2}uR)}{I_0(\tfrac{1}{2}uR)},
&& (n=0)
\\[5pt]
& A_n = \frac{2S_0 R}{u} (-1)^n \frac{I_n'(\tfrac{1}{2}uR)}{I_n(\tfrac{1}{2}uR)}.
&& (n\ge1)
\end{align}

Finally, combining the particular and homogeneous solutions ($\phi = \phi_p + \phi_h$) gives the exact analytical solution
\begin{align}\label{u-drift-solution}
\phi(r,\varphi) =
\frac{S_0R}{u} & \Bigg\{
\frac{r\cos\varphi}{R} + e^{\tfrac{1}{2}ur\cos\varphi}
\Bigg[
\frac{ I_1(\tfrac{1}{2}uR) }{ I_0(\tfrac{1}{2}uR) } \, I_0(\tfrac{1}{2}ur)
\nonumber\\[10pt]
&\!\!\!\!
+ 2 \sum_{n=1}^\infty (-1)^n \, \frac{ I'_n(\tfrac{1}{2}uR) }{ I_n(\tfrac{1}{2}uR) } I_n(\tfrac{1}{2}ur) \cos(n\varphi) \Bigg] \Bigg\}.
\end{align}
The prefactor $1/u$ does not produce a singular potential in the purely diffusive limit $u\to0$. In this limit, the leading $n=1$ contribution from the Bessel series cancels the explicit term $(r/R)\cos\varphi$ from the particular solution when the exponential prefactor is expanded for small $u$. The remaining contributions are of order $u$ or higher, so that $\phi$ remains finite as $u\to0$.

\begin{figure*}[tb]
	\centering
	\begin{subfigure}{0.30\textwidth}
		\centering        \includegraphics[width=.7\linewidth]{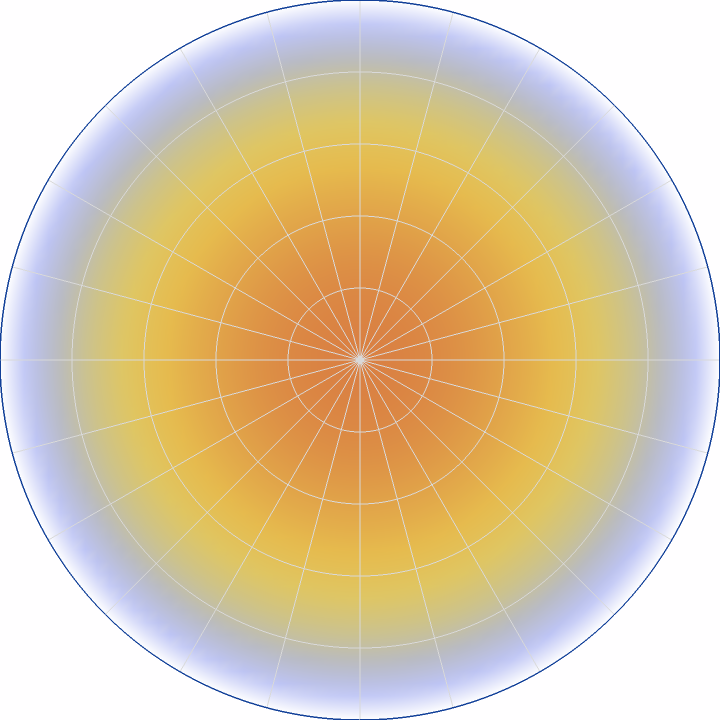}
	\end{subfigure}
	\hfill
	\begin{subfigure}{0.30\textwidth}
		\centering        \includegraphics[width=.7\linewidth]{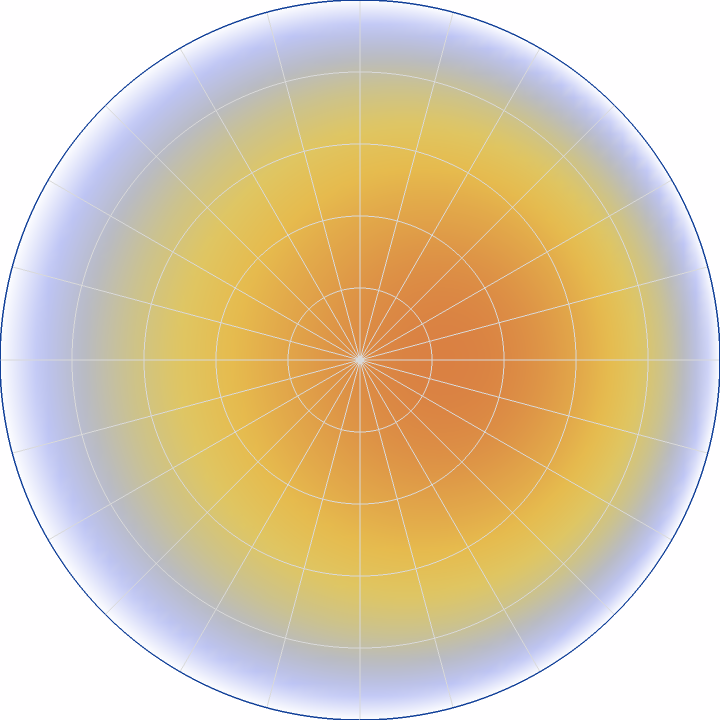}
	\end{subfigure}
	\hfill
	\begin{subfigure}{0.30\textwidth}
		\centering        \includegraphics[width=.7\linewidth]{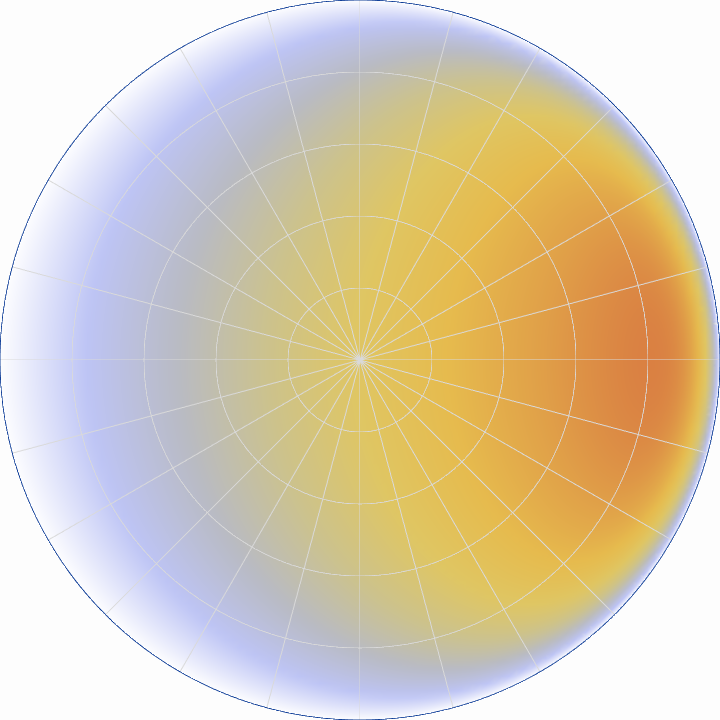}
	\end{subfigure}

	\vspace{1em}

	\begin{subfigure}{0.30\textwidth}
		\centering
		\includegraphics[width=.7\linewidth]{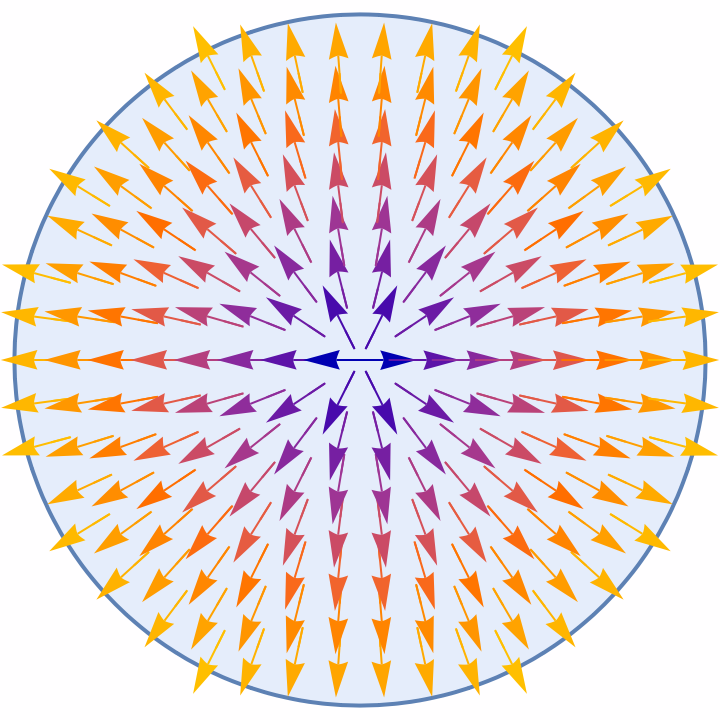}
	\end{subfigure}
	\hfill
	\begin{subfigure}{0.30\textwidth}
		\centering        \includegraphics[width=.7\linewidth]{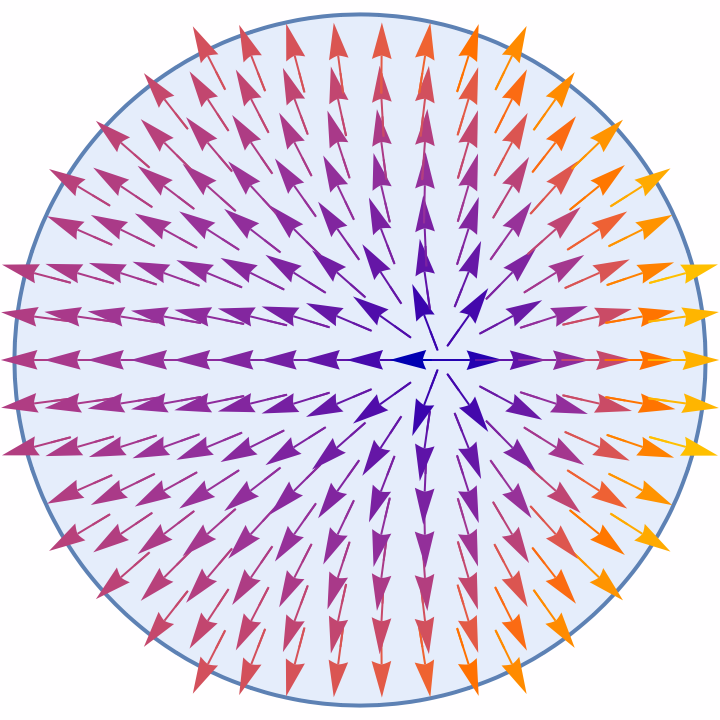}
	\end{subfigure}
	\hfill
	\begin{subfigure}{0.30\textwidth}
		\centering        \includegraphics[width=.7\linewidth]{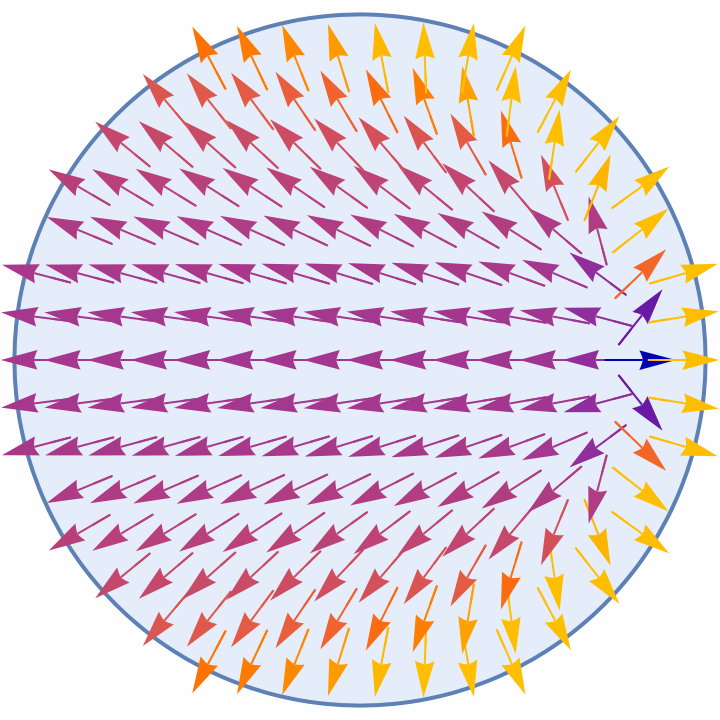}
	\end{subfigure}

	\vspace{1em}

	\begin{subfigure}{0.30\textwidth}
		\centering        \includegraphics[width=.7\linewidth]{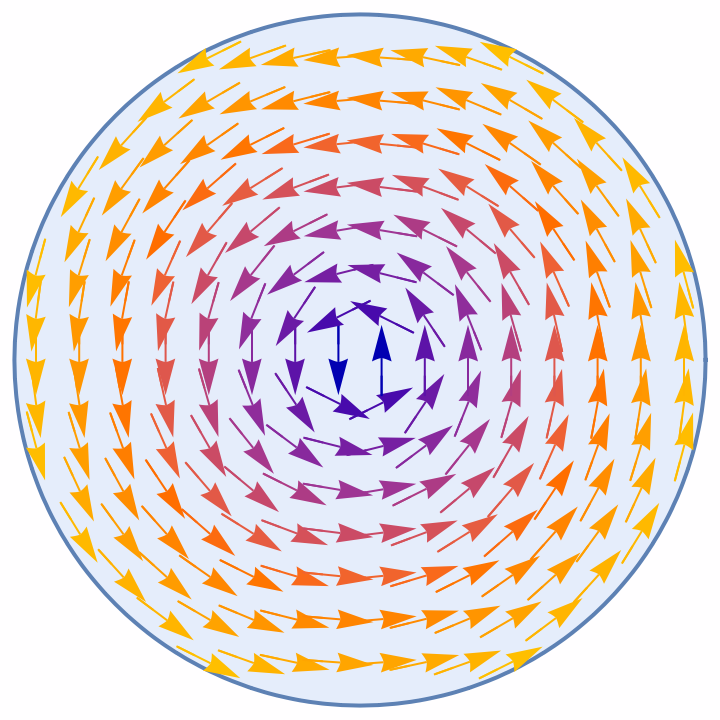}
		\caption{}
		\label{fig:uniform-DIFF}
	\end{subfigure}
	\hfill
	\begin{subfigure}{0.30\textwidth}
		\centering        \includegraphics[width=.7\linewidth]{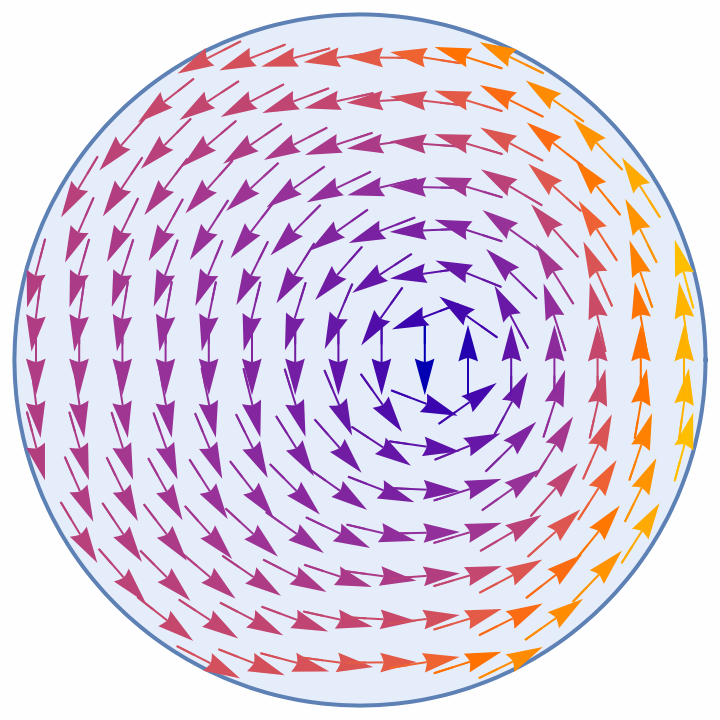}
		\caption{}
		\label{fig:uniform-TRANSITION}
	\end{subfigure}
	\hfill
	\begin{subfigure}{0.30\textwidth}
		\centering        \includegraphics[width=.7\linewidth]{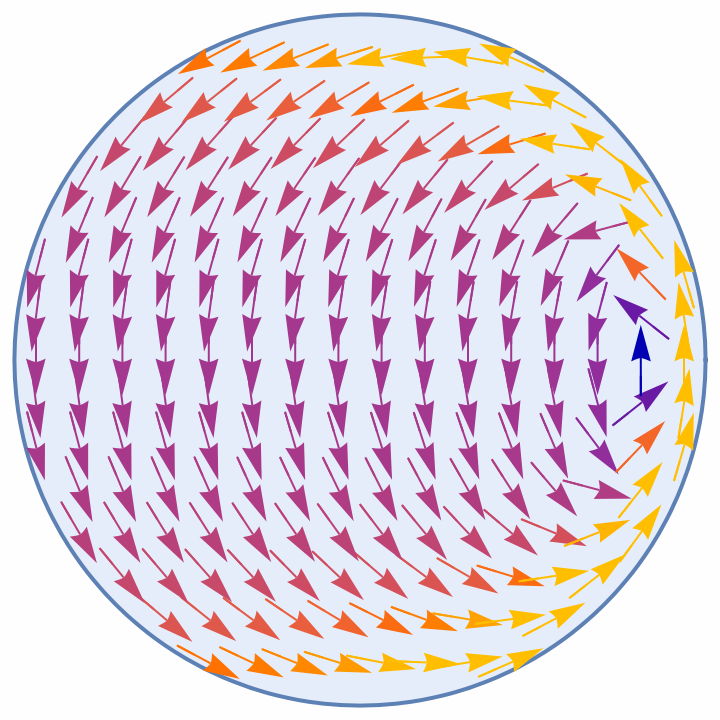}
		\caption{}
		\label{fig:uniform-ADV}
	\end{subfigure}

	\caption{Fields $\phi$ (top), $\ep$ (middle), and $\bp$ (bottom) in the circular guide for uniform drift at $\text{Pe}$ equal to (a) $0$, (b) $1$, and (c) $10$. Warmer (cooler) colors denote larger (smaller) field magnitudes.}
	\label{fig:disk-uniform}
\end{figure*}

The P\'{e}clet number $\text{Pe} = b_\bot R = uR$ appears naturally in \eqref{u-drift-solution}. Figure \ref{fig:disk-uniform} shows the corresponding $\ep$ and $\bp$ fields in the purely diffusive regime ($\text{Pe}=0$), the crossover regime ($\text{Pe}=1$), and the advection-dominated regime ($\text{Pe}=10$), for $\bu=u\hat{\mbf{x}}$ or, equivalently, $\bbot=b_\bot \hat{\mbf{y}}$. At large P\'{e}clet number, the field variations become confined to a layer of thickness
\be
\delta \sim \frac{1}{u}.
\ee
In the bulk, the electric and magnetic fields tend to align parallel and perpendicular to the drift field, respectively. The limiting forms of the solution for $\text{Pe}\ll1$ and $\text{Pe}\gg1$ are derived explicitly in the following sections.

\subsubsection{Small-P\'{e}clet-number regime}
\label{sec:uniform-drift-small-Pe}

For $\text{Pe}\ll1$, the exact solution can be expanded systematically in powers of the P\'{e}clet number to describe the departure from the purely diffusive state, as illustrated in Fig.~\ref{fig:uniform-TRANSITION}. Introducing the dimensionless radial coordinate $\rho=r/R$ and using the small-argument expansions of the modified Bessel functions  \cite{arfken},
\be
I_n(z)=\frac{(z/2)^n}{n!}  \left[ 1 + \frac{z^2}{4(n+1)} + \mathcal{O}(z^4)  \right],
\ee
and
\be
\frac{I_n'(z)}{I_n(z)} = \frac{n}{z} + \frac{z}{2(n+1)} + \mathcal{O}(z^3),
\ee
the exact solution can be expanded in powers of $\text{Pe}=uR$. Although individual terms in the Bessel series contain factors that diverge as $\text{Pe}\to0$, these divergences are canceled by the corresponding powers of $I_n(\text{Pe}\rho/2)$, leaving the full solution finite in this limit.

To second order in $\text{Pe}$, the resulting expansion is
\begin{align}
\frac{\phi(r,\varphi)}{S_0R^2} =
& \frac{1-\rho^2}{4} + \text{Pe}\, \frac{\rho(1-\rho^2)}{16} \cos\varphi
\nonumber\\[5pt]
& - \text{Pe}^2  \left[ \frac{(1-\rho^2)^2}{128} - \frac{\rho^2(1-\rho^2)}{192}\cos(2\varphi)  \right]
\nonumber\\[5pt]
& + \mathcal{O}(\text{Pe}^3).
\end{align}
Equivalently, in dimensional variables,
\begin{align}\label{u-drift-small-Pe-solution}
\phi(r,\varphi) =
& \frac{S_0}{4}(R^2-r^2) + \frac{S_0u}{16}\,r(R^2-r^2)\cos\varphi
\nonumber\\[5pt]
& - S_0 u^2  \left[ \frac{(R^2-r^2)^2}{128} - \frac{r^2(R^2-r^2)}{192}\cos(2\varphi)  \right]
\nonumber\\[5pt]
& + \mathcal{O}(u^3R^5).
\end{align}
The zeroth-order term recovers the purely diffusive solution \eqref{diffusion-disc-solution} and is rotationally symmetric. The leading advective correction appears at first order in $\text{Pe}$ and has a purely dipolar angular dependence, proportional to $x=r\cos\varphi$. Thus, rotational symmetry is broken already at $\mathcal{O}(\text{Pe})$ by the directional bias along the $x$-axis. At second order, an additional rotationally symmetric correction appears, together with a quadrupolar contribution proportional to $\cos(2\varphi)$. Higher powers of $\text{Pe}$ generate progressively higher angular harmonics. In particular, the $n$th angular harmonic first appears at order $\text{Pe}^n$, reflecting the progressive development of anisotropy induced by the uniform drift.

\subsubsection{Large-P\'{e}clet-number regime}
\label{sec:uniform-drift-large-Pe}

For $\text{Pe}\gg1$, the field profiles in Fig.~\ref{fig:uniform-ADV} suggest that advection dominates diffusion in the bulk, whereas diffusion remains essential within a thin boundary layer where the fields vary rapidly. We analyze this behavior using the method of matched asymptotic expansions, following Refs.~\cite{VanDyke1975,Gie2018}.

\begin{figure}[htb]
	\centering
	\includegraphics[width=.6\linewidth]{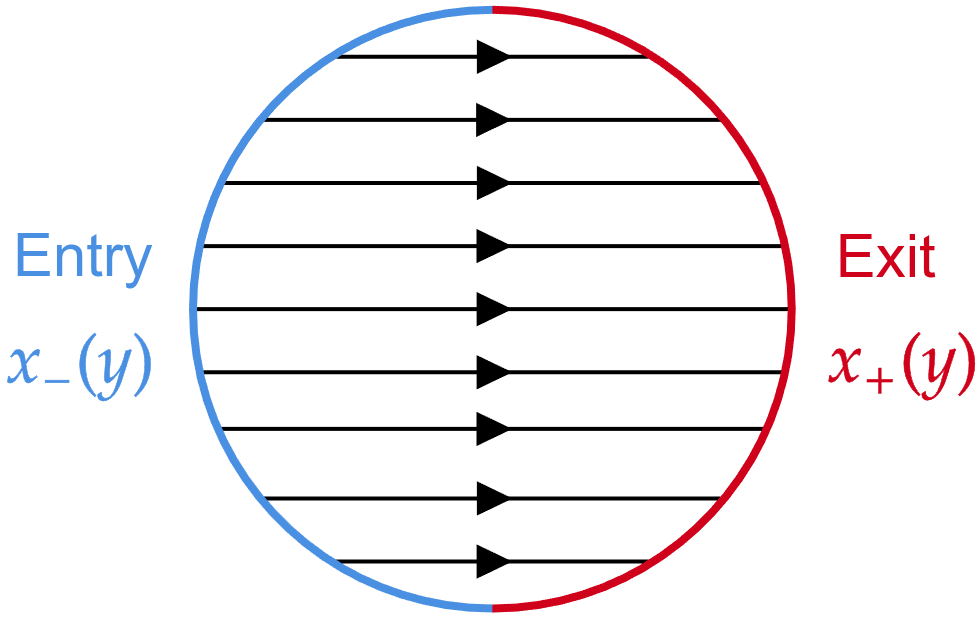}
	\caption{Identification of entry (blue) and exit (red) boundaries of uniform drift lines in the cross section of a circular guide.}
	\label{fig:stream-uniform}
\end{figure}

The boundary of the waveguide cross section is described by
\be
x_\pm(y) = \pm \sqrt{R^2-y^2}.
\ee
Taking the drift direction along the $x$ axis, $\bu = u\hat{\bx}$ with $u > 0$, drift lines enter the domain at $x_-(y)$ and leave it at $x_+(y)$, as illustrated in Fig.~\ref{fig:stream-uniform}. The boundary layer forms at the exit boundary $x_+(y)$, as shown in Fig.~\ref{fig:uniform-ADV}. Accordingly, we decompose the asymptotic solution as
\be\label{u-drift-large-Pe-asymptotic}
\phi=\phi_\text{bulk}+\phi_\text{layer},
\ee
where $\phi_\text{bulk}$ is the smooth bulk solution, away from the accumulation layer, and $\phi_\text{layer}$ is the localized correction near the exit boundary $x_+$.

In the bulk, the diffusion term in Eq.~\eqref{advection-diffusion} is subleading, and the advection--diffusion equation therefore reduces to
\be
u \partial_x \phi_\text{bulk}  \simeq  S_0.
\ee
Imposing the Dirichlet condition at the entry boundary, $\phi_\text{bulk}(x_-,y)=0$, yields
\be\label{uniform-drif-large-Pe-bulk}
\phi_\text{bulk}(x,y)  \simeq  \frac{S_0}{u}  \left( x + \sqrt{R^2-y^2}  \right).
\ee
This expression gives the leading-order solution for $x<0$. However, it does not satisfy the boundary condition for $x>0$: at the exit boundary, the bulk solution takes the value
\be
\phi_\text{bulk}(x_+,y)  \simeq  \frac{2S_0}{u} \sqrt{R^2-y^2},
\ee
which is nonzero, whereas the Dirichlet boundary condition requires $\phi=0$ on $r=R$. Diffusion must therefore be retained in a thin layer adjacent to $x_+(y)$, where $\phi_\text{layer}$ compensates for the nonzero boundary value of $\phi_\text{bulk}$.

Substituting the asymptotic decomposition \eqref{u-drift-large-Pe-asymptotic} into the advection--diffusion equation, we find that $\phi_\text{layer}$ satisfies, near the exit boundary,
\be\label{u-drift-large-Pe-layer-eq}
\nabla^2\phi_\text{layer} - u \partial_x \phi_\text{layer}   \simeq   0,
\ee
subject to the boundary condition
\be\label{u-drift-large-Pe-BC}
\phi_\text{layer} (x_+,y) \simeq  -\phi_\text{bulk}(x_+,y).
\ee
It is convenient to introduce polar coordinates and the inward normal coordinate $s$, which measures the distance from the circular boundary, $s = R - r$. Since $\partial_r=-\partial_s$ and $\partial_x \simeq \cos\varphi \partial_r$ near the waveguide wall, the normal advection term $-u\partial_x\phi_\text{layer}$ becomes $+u\cos\varphi\partial_s\phi_\text{layer}$. Here, $\bu\cdot\hat{\br}=u\cos\varphi>0$ is the normal component of the drift at the exit boundary. Near $r = R$, the leading-order form of Eq.~\eqref{u-drift-large-Pe-layer-eq} is therefore
\be
\partial_s^2 \phi_\text{layer} + u \cos\varphi \partial_s \phi_\text{layer}  \simeq  0,
\ee
Within the layer, the leading-order equation contains only normal diffusion and normal advection. Tangential diffusion, the $1/r$ term in the Laplacian, and tangential advection are smaller by factors of order $1/\text{Pe}$, $1/\text{Pe}$, and $1/\text{Pe} \cos^2\varphi$, respectively. The solution that decays from the exit boundary into the interior and satisfies the boundary condition \eqref{u-drift-large-Pe-BC} is
\be\label{u-drift-large-Pe-layer}
\phi_\text{layer}(r,\varphi)  \simeq  -\frac{2S_0R\cos\varphi}{u} e^{-u(R-r) \cos\varphi}.
\ee

Combining Eqs.~\eqref{uniform-drif-large-Pe-bulk} and \eqref{u-drift-large-Pe-layer}, we obtain the composite asymptotic solution for $x>0$:
\be
\phi(r,\varphi)  \simeq \frac{S_0}{u} \Bigg[ x + \sqrt{R^2-y^2} - 2R \cos\varphi e^{-u(R-r) \cos\varphi} \Bigg].
\ee
At the exit boundary $x_+(y)$, the composite solution satisfies the Dirichlet condition to leading order. Away from the waveguide wall, the exponential correction becomes negligible and the bulk solution is recovered. The characteristic boundary-layer thickness is therefore
\be
\delta(\varphi)
\sim \frac{1}{u \cos\varphi}.
\ee
Thus, the thickness measured normal to the circular boundary depends on the local angle $\varphi$ between the outward normal $\hat\br$ and the drift direction $\hat\bu$. Along the drift direction, this reduces to
\be
\delta\sim\frac{1}{u} = \frac{R}{\text{Pe}}.
\ee
The divergence of the estimate $\delta(\varphi)$ as $\cos\varphi\to0$ signals the breakdown of the local boundary-layer approximation near the points where the drift is tangent to the waveguide wall.

The exit boundary layer is also associated with a strongly localized electric field and a corresponding transverse magnetic field. In the bulk, the potential varies over the characteristic length $R$, resulting in a field of magnitude
\be
|\ep_\text{bulk}|\sim\frac{S_0}{u} = \frac{S_0 R}{\text{Pe}}.
\ee
Near the exit boundary, the Dirichlet condition requires the bulk potential to decrease to zero across a thin layer of thickness $\delta\sim1/u$ along the drift direction. Since the potential variation across this layer remains of order $S_0R/u$, the associated field scales as
\be
|\ep_\text{layer}|\sim S_0R.
\ee
The resulting amplification relative to the bulk field is
\be
\frac{|\ep_\text{layer}|}{|\ep_\text{bulk}|}
\sim uR
=\text{Pe}.
\ee
Thus, the large-P\'{e}clet-number limit produces a pronounced separation between the field magnitudes in the bulk and in the boundary layer. While the bulk field decreases as $\text{Pe}^{-1}$, the field within the exit layer remains of order $S_0R$, reflecting the steep spatial variation required to satisfy the boundary condition.

\subsection{Nonuniform drift and the Helmholtz reduction}

The reduction to the modified Helmholtz equation \eqref{helmholtz} in Sec.~\ref{sec:uniform-drift} can be generalized, under certain conditions, to nonuniform irrotational drift fields, $\bu = \nabla f$. In this case, introducing the transformation
\be
\phi=e^{f(\br)/2}\psi
\ee
eliminates the first-order derivative and yields
\be
\nabla^2\psi - U(\br)\psi = -S_0e^{-f(\br)/2},
\ee
where
\be
U(\br) = \frac{\bu^2}{4} - \frac{\nabla \cdot \bu}{2}.
\ee
Thus, for an irrotational drift field, the first-order advection term can be absorbed into an effective scalar field $U(\br)$. When
\be
\bu^2-2\nabla\cdot\bu = 4U = \text{const},
\ee
the differential operator becomes the constant-coefficient modified Helmholtz operator considered in the uniform-drift case, although the transformed source term generally remains spatially dependent.

This reduction therefore requires two distinct conditions. Irrotationality, $\nabla\times\bu=0$, is necessary for the scalar transformation above to exist, whereas the additional condition that $U(\br)$ be constant is required to obtain a constant-coefficient modified Helmholtz operator. Consequently, nonzero vorticity, $\nabla\times\bu\neq0$, prevents this scalar reduction. According to Eq.~\eqref{vorticity}, this obstruction occurs whenever the in-plane axion background field is non-harmonic, $\nabla^2\theta_{xy} \neq 0$.

The nonuniform drift profiles considered below generally do not satisfy these conditions and therefore require a direct treatment of the advection--diffusion equation.

\subsection{Rigid rotational drift}
\label{sec:rotational-drift}

In Sec.~\ref{sec:uniform-drift}, we considered TEM modes in a circular waveguide with rotationally symmetric boundary conditions. Although the geometry and boundary conditions are rotationally symmetric, a uniform drift breaks this symmetry, producing anisotropic field profiles and accumulation near preferential portions of the waveguide wall. Here, we consider rigid rotational drift, whose drift field is axisymmetric and has a well-defined sense of circulation.

The rigid rotational drift about the $z$ axis is characterized by
\be
\bu = (\Omega\z) \times \br =  \Omega r \hat{\bs{\varphi}}
\ee
where $\Omega$ plays a role analogous to an angular velocity, although it has dimensions of inverse length squared. Through $\bu = \bbot \times \z$, this drift corresponds to the nonuniform transverse anisotropy
\be
\bbot = - \Omega \br
\ee
which originates from the in-plane axion background
\be\label{axion-rotational-drift}
\theta_{xy}^{(\Omega)} = \frac{1}{2}\Omega(x^2+y^2).
\ee
Unlike uniform drift, the present drift field is spatially nonuniform and has nonzero vorticity, $\nabla\times\bu = \nabla^2\theta_{xy}^{(\Omega)}\z = 2\Omega\z$. Consequently, the advection--diffusion equation cannot be reduced to a modified Helmholtz equation, as in the uniform-drift case.

For this drift field, the advection--diffusion equation takes the form
\be\label{solid-body-rotation}
\nabla^2\phi - \Omega \frac{\partial\phi}{\partial\varphi} = - S_0.
\ee
In a circular waveguide, the rotationally symmetric Dirichlet boundary condition $\phi(R,\varphi)=0$ leads to the same solution as in the pure-diffusion case, Eq.~\eqref{diffusion-disc-solution}. Indeed, rotationally symmetric boundary conditions admit a radial solution $\phi=\phi(r)$, for which $\partial_\varphi\phi=0$ and the advective term vanishes identically. Thus, rigid rotational drift does not affect the solution in this geometry. To isolate the effects of rotational advection, we therefore consider boundary conditions that break rotational symmetry. In Sec.~\ref{sec:combined-drift}, we further examine the interplay between rigid rotational drift and uniform drift.

A simple geometry in which rigid rotational drift produces nontrivial effects is an elliptical cylindrical guide. We impose the Dirichlet boundary condition
\be\label{BC-ellipse}
\phi(r,\varphi)|_{R(\varphi)} = 0,
\ee
where
\be\label{elliptical-radius}
R(\varphi) = \frac{ab}{\sqrt{a^ 2\sin^2\varphi + b^2 \cos^2\varphi}},
\ee
and $a$ and $b$ are the major and minor semiaxes of the elliptical waveguide cross-sectional boundary, respectively.

\begin{figure*}[tb]
    \centering
    \begin{subfigure}{0.30\textwidth}
        \centering        \includegraphics[width=\linewidth]{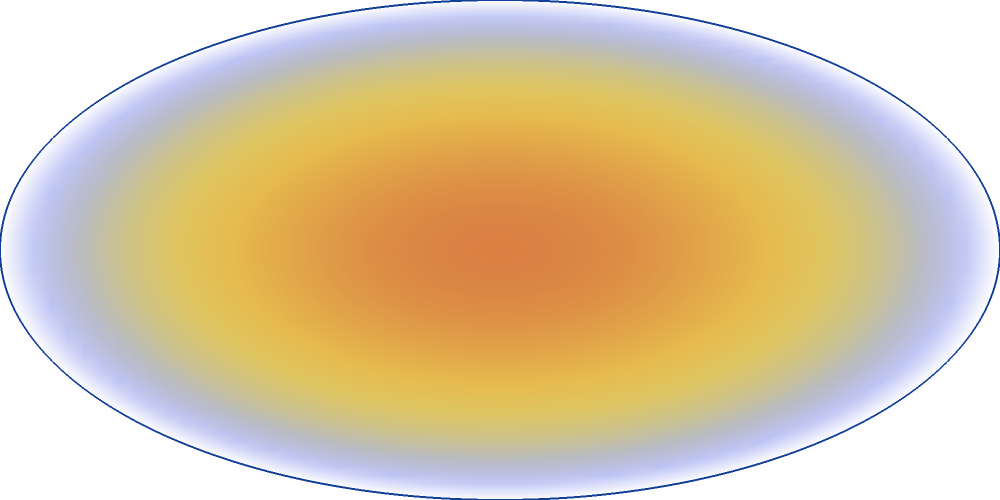}
    \end{subfigure}
        \hfill
    \begin{subfigure}{0.30\textwidth}
        \centering        \includegraphics[width=\linewidth]{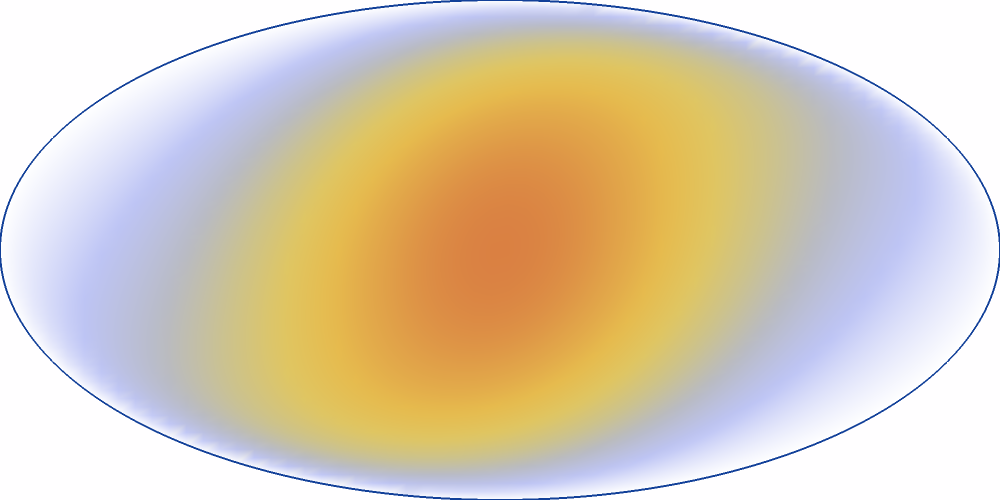}
    \end{subfigure}
        \hfill
    \begin{subfigure}{0.30\textwidth}
        \centering        \includegraphics[width=\linewidth]{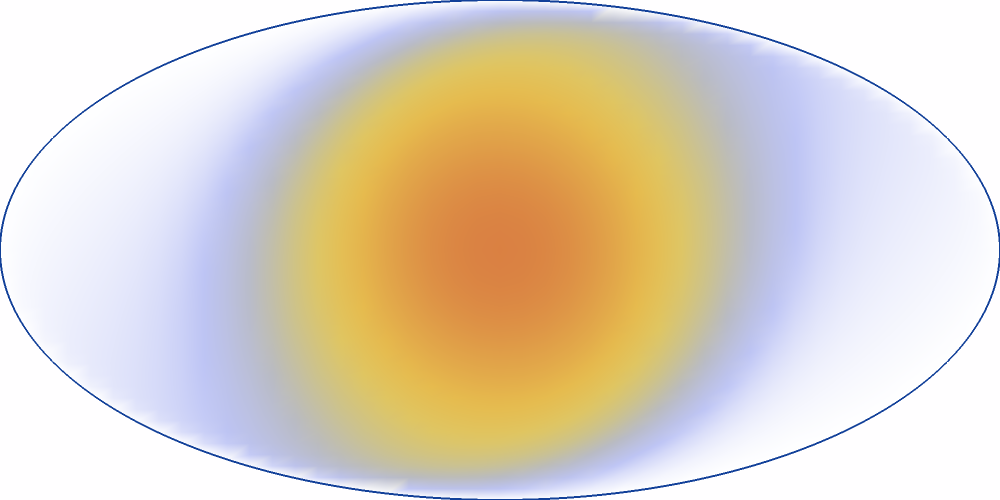}
    \end{subfigure}

    \vspace{1em}

    \begin{subfigure}{0.30\textwidth}
        \centering
        \includegraphics[width=\linewidth]{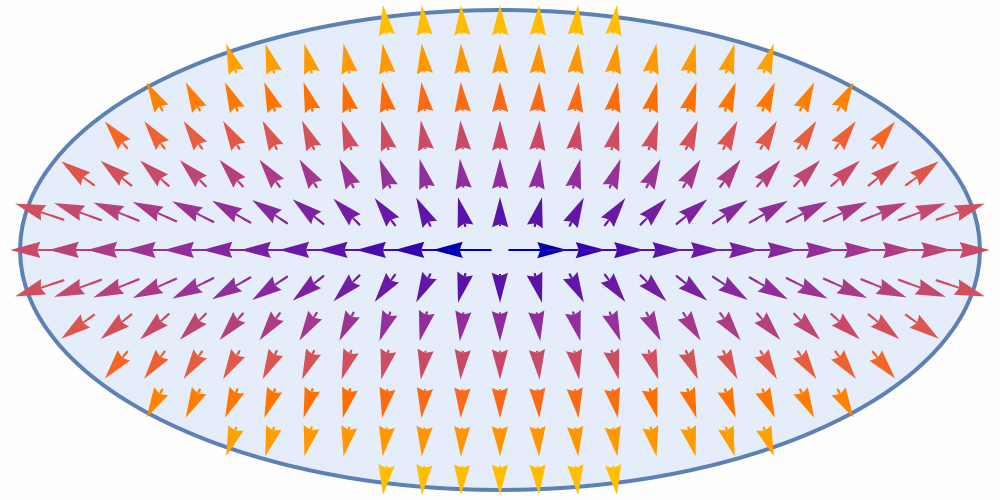}
    \end{subfigure}
    \hfill
    \begin{subfigure}{0.30\textwidth}
        \centering        \includegraphics[width=\linewidth]{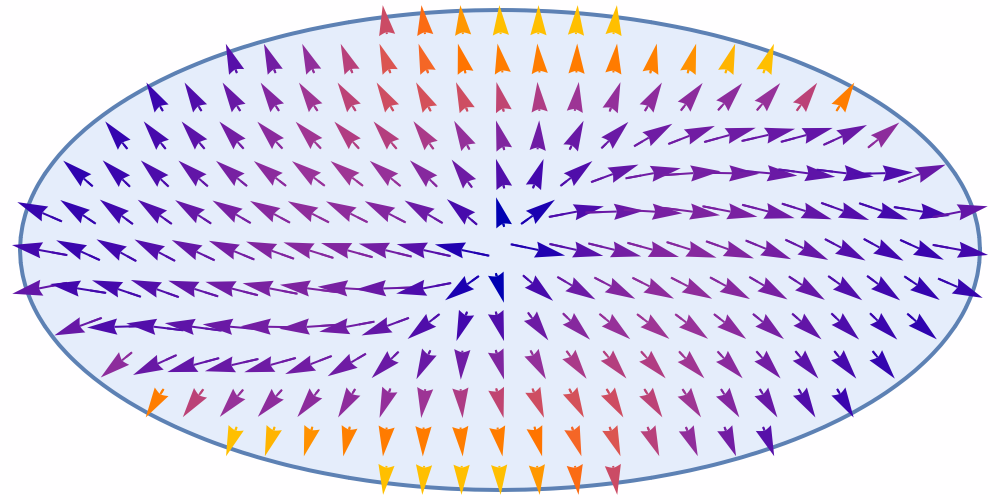}
    \end{subfigure}
    \hfill
    \begin{subfigure}{0.30\textwidth}
        \centering        \includegraphics[width=\linewidth]{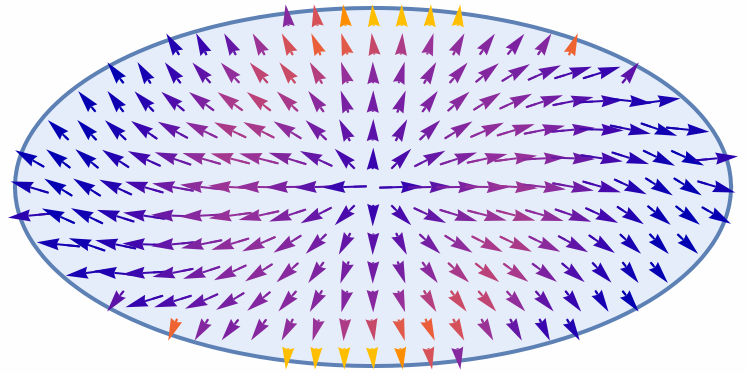}
    \end{subfigure}

    \vspace{1em}

    \begin{subfigure}{0.30\textwidth}
        \centering        \includegraphics[width=\linewidth]{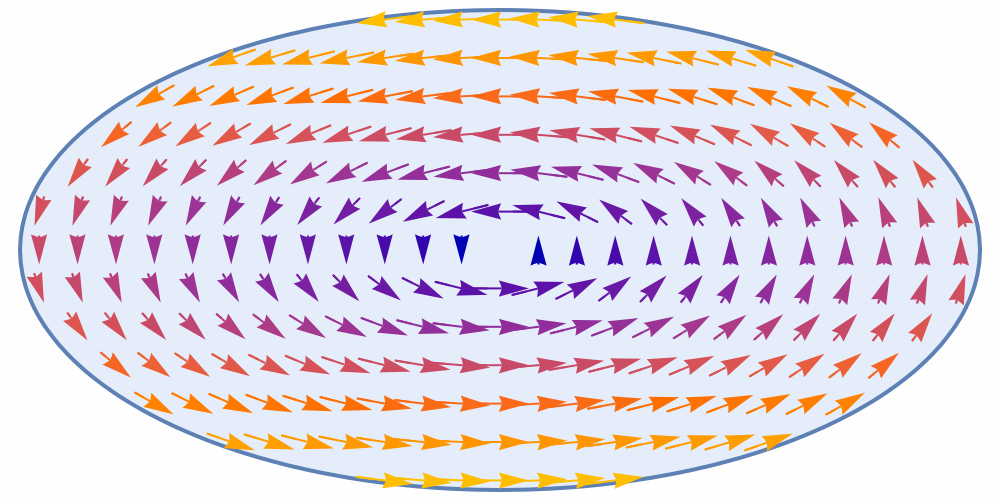}
        \caption{}
        \label{fig:ellipse-DIFF}
    \end{subfigure}
    \hfill
    \begin{subfigure}{0.30\textwidth}
        \centering        \includegraphics[width=\linewidth]{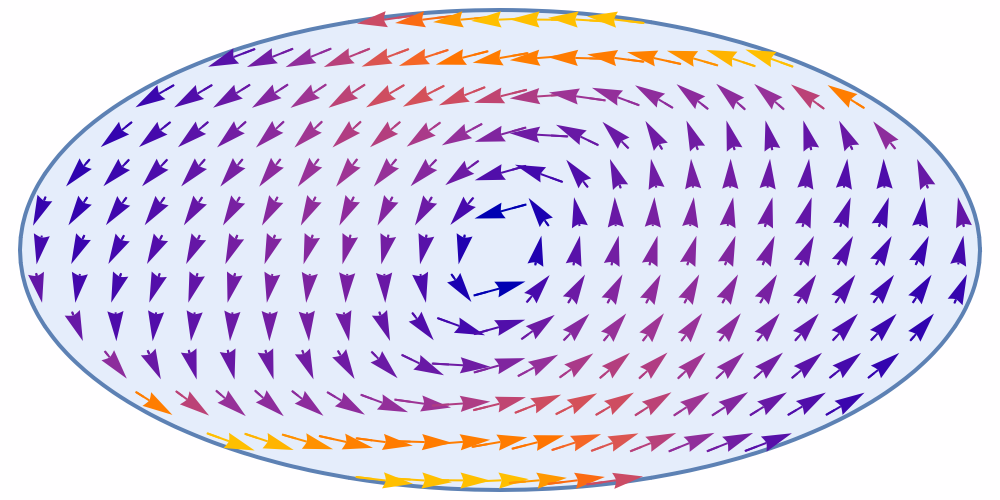}
        \caption{}
        \label{fig:ellipse-TRANSITION}
    \end{subfigure}
    \hfill
    \begin{subfigure}{0.30\textwidth}
        \centering        \includegraphics[width=\linewidth]{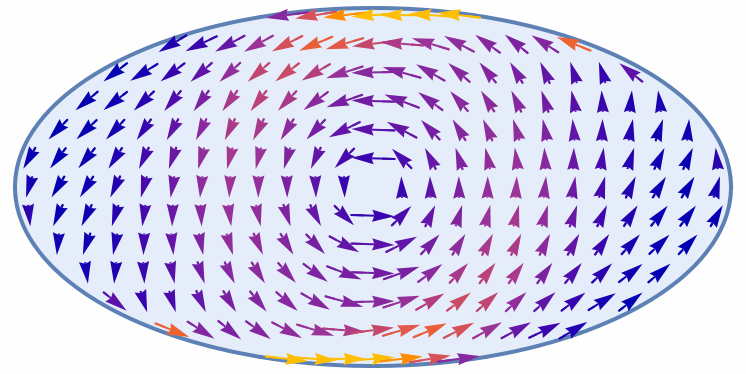}
        \caption{}
        \label{fig:ellipse-ADV}
    \end{subfigure}

    \caption{Fields $\phi$ (top), $\ep$ (middle), and $\bp$ (bottom) in the elliptical cylindrical guide for rigid rotational drift at $\text{Pe}_\Omega$ equal to (a) $0$, (b) $40$, and (c) $150$. Warmer (cooler) colors denote larger (smaller) field magnitudes.}
    \label{fig:ellipse}
\end{figure*}

The pure-diffusion solution satisfying the boundary condition at the elliptical waveguide boundary is
\be\label{ellipse-pure-diffusion}
\phi(x,y) = \frac{S_0}{2} \frac{a^2b^2}{a^2+b^2}  \left( 1 - \frac{x^2}{a^2} - \frac{y^2}{b^2}  \right),
\ee
which can be readily verified to satisfy the advection--diffusion equation for $\Omega=0$. The corresponding profiles of $\phi$, $\ep$, and $\bp$ are shown in Fig.~\ref{fig:ellipse-DIFF}.

For $\Omega \neq 0$, the solution exhibits nontrivial effects of rotational drift and can be written as the sum of a particular solution, $-S_0 r^2/4$, and a Fourier expansion,
\be
\phi(r,\varphi) = - \frac{S_0 r^2}{4} + \sum_{m=-\infty}^\infty A_m f_m(r) e^{im\varphi}.
\ee
Substituting the Fourier expansion into Eq.~\eqref{solid-body-rotation} separates the angular and radial variables. For each angular mode $m$, the radial function satisfies
\be
r^2 f'' + r f' - (k_m^2 r^2 + m^2) f = 0,
\ee
where $k_m^2 = im\Omega$ and $f'\equiv df/dr$. The solutions that are regular at the origin are the modified Bessel functions of the first kind, $I_m(k_m r)$. Since both the order and the argument of the Bessel function depend on the angular mode index $m$, the radial profile of each Fourier component is mode dependent. In particular, $k_m\propto \sqrt{m\Omega}$ sets an advection-induced radial length scale of order $|k_m|^{-1}$, while the order $m$ determines the additional near-origin behavior. The solution can therefore be expressed as
\be
\phi(r,\varphi) = -\frac{S_0 r^2}{4} + \sum_{m=-\infty}^\infty A_m I_m(k_m r) e^{im\varphi}.
\ee
To ensure that $\phi$ is real-valued, we take $\sqrt{i}$ on the principal branch ($-\pi < \text{arg }z < \pi$), such that $\sqrt{i}=(\sqrt{-i})^*$. Since $I_n(z^*) = I_n^*(z)$, the reality condition is satisfied by choosing $A_{-m}^*=A_m$. Accordingly,
\begin{align}\label{ellipse-phi-unknowns}
\phi(r,\varphi) =
& -\frac{S_0 r^2}{4} + A_0
+ \sum_{m=1}^\infty \alpha_m \text{Re} \left\{ I_m(k_m r) e^{im\varphi}  \right\}
\nonumber\\[5pt]
& + \sum_{m=1}^\infty \beta_m \text{Im} \left\{ I_m(k_m r) e^{im\varphi}  \right\},
\end{align}
where $A_0$ is real and the real-valued coefficient pairs $(\alpha_m,\beta_m)$ are defined in terms of the complex coefficients $A_m$ by $A_m=\tfrac{1}{2}(\alpha_m-i\beta_m)$.

Because the elliptical boundary is described by a $\varphi$-dependent radius $R(\varphi)$ in the polar coordinates used for the expansion, the coefficients cannot be determined analytically from the boundary condition \eqref{BC-ellipse}. Moreover, the representation above becomes singular at $\Omega=0$, since the basis functions with nonzero $m$ vanish in this limit. The case $\Omega=0$ is therefore treated separately using the pure-diffusion solution discussed above. For $\Omega>0$, we determine $A_0$, $\alpha_m$, and $\beta_m$ numerically by imposing the Dirichlet boundary condition \eqref{BC-ellipse} at a set of collocation points. We choose $N_\varphi$ equally spaced values of the polar angle $\varphi$ along the elliptical boundary and evaluate the truncated expansion at the corresponding points $r=R(\varphi)$. For a truncation at angular mode $M$, imposing $\phi(R(\varphi),\varphi)=0$ at each collocation point gives a linear equation for the $2M+1$ unknown real coefficients $A_0$, $\alpha_m$, and $\beta_m$ ($1\le m \le M$). The real and imaginary parts of the basis functions $I_m(k_m r)e^{im\varphi}$ are treated as separate basis functions, together with the constant term $A_0$. Choosing $N_\varphi > 2M+1$ therefore yields an overdetermined linear system.

Rather than enforcing the boundary condition exactly at a finite set of points, which is generally not possible with a truncated expansion, we determine the coefficients by linear least squares. Specifically, the coefficients are chosen to minimize the squared Euclidean norm of the residual vector, whose components are $\mathcal{R}_j\equiv\phi(R(\varphi_j),\varphi_j)$ at the collocation points. The collocation parameters $N_\varphi$ and $M$ are increased until the solution and the observables of interest converge to the desired accuracy. For the plots in Fig.~\ref{fig:ellipse}, we use $N_\varphi=300$ collocation points and retain angular modes up to $M=35$. Increasing $N_\varphi$ further produces no appreciable changes in the resulting solution or in the boundary-condition residual, whereas increasing $M$ systematically reduces the residual.

For a characteristic transverse length scale $L=a$, where $a$ is the major semiaxis, the relative strength of rotational advection and diffusion is characterized by the rotational P\'{e}clet number
\be
\text{Pe}_\Omega = \Omega a^2.
\ee
The limits $\text{Pe}_\Omega\ll 1$ and $\text{Pe}_\Omega\gg 1$ correspond, respectively, to diffusion-dominated and advection-dominated transport. Figure \ref{fig:ellipse} shows the corresponding $\phi$, $\ep$ and $\bp$ profiles in the purely diffusive regime ($\text{Pe}_\Omega=0$), an advection-dominated regime ($\text{Pe}_\Omega=40$), and a regime of even stronger advection ($\text{Pe}_\Omega=150$). At large P\'{e}clet number, the fields in the bulk tend toward a rotationally symmetric profile, despite the elliptical diffusive pattern imposed by the waveguide boundary. The resulting distortions are confined to a boundary layer of thickness
\be
\delta \sim \frac{1}{\Omega a}.
\ee
The limiting forms of the scalar potential $\phi$ for $\text{Pe}\ll1$ and $\text{Pe}\gg1$ are derived explicitly in the following sections.

\subsubsection{Small-P\'{e}clet-number regime}
\label{sec:rotational-drift-small-Pe}

In the weak-advection regime, $\text{Pe}_\Omega\ll 1$, rotational advection can be treated perturbatively. We therefore expand the solution of Eq.~\eqref{solid-body-rotation} as
\be
\phi(r,\varphi) = \sum_{n=0}^\infty (\text{Pe}_\Omega)^n \phi^{(n)}(r,\varphi).
\ee
Substituting this expansion into the advection--diffusion equation yields the hierarchy
\be
\nabla^2\phi^{(0)} = - S_0,
\qquad
\nabla^2\phi^{(n)} = \frac{1}{a^2} \frac{\partial \phi^{(n-1)}}{\partial \varphi},
\quad n\geq 1,
\ee
with $\phi^{(n)}|_{R(\varphi)}=0$ at every order. Thus, the zeroth-order solution describes pure diffusion, while each successive correction is obtained by solving a Poisson equation with a source determined by the preceding order.

For the elliptical guide, the zeroth-order solution is given by Eq.~\eqref{ellipse-pure-diffusion}:
\be
\phi^{(0)}(x,y) = \frac{S_0}{2} \frac{a^2b^2}{a^2+b^2}  \left( 1 - \frac{x^2}{a^2} - \frac{y^2}{b^2}  \right).
\ee
The corresponding transverse field profiles are shown in Fig.~\ref{fig:ellipse-DIFF}. This expression satisfies both $\nabla^2\phi^{(0)}=-S_0$ and the Dirichlet boundary condition on the ellipse. In polar coordinates, its angular dependence contains only the azimuthally symmetric component and the $m=2$ harmonic,
\be
\phi^{(0)}(r,\varphi) = C^{(0)}_0(r) + C^{(0)}_2(r)\cos2\varphi,
\ee
where
\be
C^{(0)}_0(r) = \frac{S_0}{2} \frac{a^2b^2}{a^2+b^2} - \frac{S_0 r^2}{4},
\ee
\be
C^{(0)}_2(r) = \frac{S_0 r^2}{4} \frac{a^2-b^2}{a^2+b^2}.
\ee
The anisotropy at zero rotational drift is thus entirely determined by the elliptic geometry. Although the angular decomposition contains a quadrupolar ($m=2$) contribution, its superposition with the azimuthally symmetric ($m=0$) component yields an anisotropic parabolic profile rather than a four-lobed spatial structure. The quadrupolar character is therefore encoded in the angular dependence of the $m=2$ harmonic, which quantifies the leading-order deviation from azimuthal symmetry induced by the elliptic waveguide boundary.

The first-order correction satisfies
\be
\nabla^2\phi^{(1)} = \frac{1}{a^2} \frac{\partial \phi^{(0)}}{\partial \varphi} = - \frac{S_0}{a^2}
\frac{a^2-b^2}{a^2+b^2} xy.
\ee
The corresponding solution is
\be
\phi^{(1)}(x,y) = \frac{1}{3a^2} \frac{a^2 - b^2}{a^2 + b^2} xy\phi^{(0)}(x,y).
\ee
Since $\phi^{(0)}(x,y)$ vanishes at the waveguide wall, the boundary condition is satisfied separately at each perturbative order. The first-order correction describes how weak rotational advection modifies the diffusion-dominated profile. Whereas $\phi^{(0)}$ is invariant under reflections about both principal axes, $\phi^{(1)}$ changes sign under either reflection. In polar coordinates, the correction therefore introduces sine harmonics with $m=2$ and $m=4$,
\be
\phi^{(1)}(r,\varphi) =
 C^{(1)}_2(r) \sin2\varphi + C^{(1)}_4(r)\sin4\varphi,
\ee
where
\be
C_m^{(1)} = \frac{1}{3ma^2} \frac{a^2 - b^2}{a^2 + b^2} r^2 C^{(0)}_{m-2}(r).
\qquad (m=2\text{ or }4)
\ee
Its amplitude is proportional to the ellipticity, as measured by $(a^2-b^2)/(a^2+b^2)$. In the circular limit $a=b$, the correction vanishes. Indeed, the radial diffusion solution remains exact for arbitrary $\Omega$ in a circular domain. Thus, the leading effect of weak rotational drift in an elliptical guide is not a uniform change in the field amplitude, but rather a reflection-odd distortion of the spatial profile. An exaggerated illustration of this behavior is shown in Fig.~\ref{fig:ellipse-TRANSITION}.

\subsubsection{Large-P\'{e}clet-number regime}
\label{sec:rotational-drift-large-Pe}

We now consider the regime of strong rotational advection, characterized by $\text{Pe}_\Omega \gg1$. The field profiles in this regime are illustrated in Figs.~\ref{fig:ellipse-TRANSITION}-\ref{fig:ellipse-ADV}. Rotational advection dominates diffusion in the bulk, leading to an approximately rotationally symmetric configuration, while diffusion is expected to dominate within thin boundary layers, where the fields must vary rapidly to satisfy the boundary conditions.

\begin{figure}[htb]
	\centering
	\includegraphics[width=.7\linewidth]{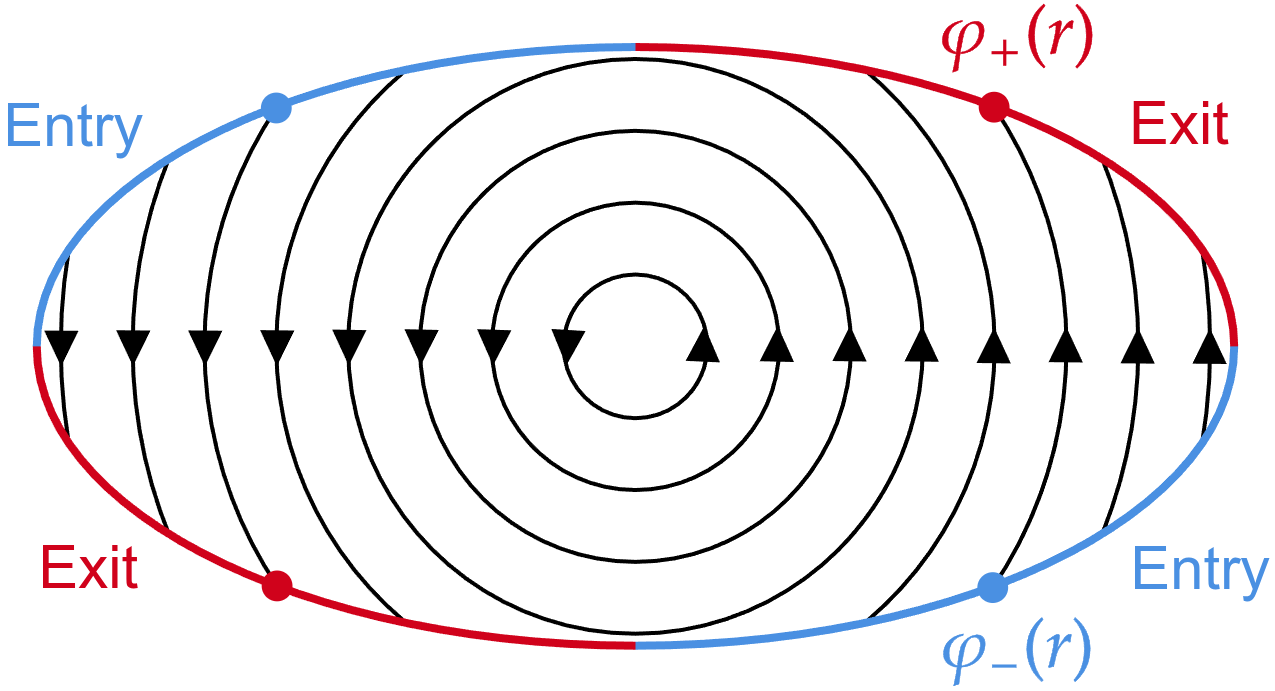}
	\caption{Identification of entry (blue) and exit (red) boundaries of rotational drift lines in the cross section of an elliptical guide.}
	\label{fig:stream-rotation}
\end{figure}

It is convenient to divide the analysis into two regions, distinguished by the circular drift lines: an outer region $r>b$, where the drift lines intersect the waveguide boundary, and a core region $r<b$, where the drift lines are closed. The drift lines in the waveguide cross section are depicted in Fig.~\ref{fig:stream-rotation}. We first determine the potential in the outer region and then match it to the potential in the core region.

The elliptical boundary is described by Eq.~\eqref{elliptical-radius}. For $\Omega>0$, the rigid rotational drift $\bu = \Omega r \hat{\bs{\varphi}}$ is directed toward increasing $\varphi$. A circular drift line $r=\text{const}$ with $b<r<a$ intersects the elliptical boundary at four points, and the portions of the drift line lying inside the ellipse form two disjoint arcs. Let $\varphi_-(r)$ and $\varphi_+(r)$ denote the entry and exit endpoints of one such arc, respectively, so that the drift enters the domain at $\varphi_-$ and leaves it at $\varphi_+$, as illustrated in Fig.~\ref{fig:stream-rotation}. The boundary layer forms at the exit endpoint $\varphi_+(r)$, as shown in Fig.~\ref{fig:ellipse-ADV}.

Following the approach used in Sec.~\ref{sec:uniform-drift-large-Pe}, we consider the asymptotic decomposition
\be
\phi_\text{outer} = \phi_\text{bulk} + \phi_\text{layer}.
\ee
In the bulk, the diffusion term in the advection--diffusion equation \eqref{solid-body-rotation} is subleading, so the equation reduces to
\be\label{rotational-drift-large-Pe-bulk-eq}
\partial_\varphi \phi_\text{bulk}  \simeq  \frac{S_0}{\Omega}.
\ee
Imposing the Dirichlet condition at the entry endpoint, $\phi_\text{bulk}(\varphi_-)=0$, gives
\be
\label{rotational-drift-large-Pe-bulk}
\phi_\text{bulk}(r,\varphi) \simeq  \frac{S_0}{\Omega} \left[ \varphi-\varphi_-(r) \right].
\ee
This expression gives the leading-order solution along the arc. However, at the exit endpoint, it takes the nonzero value
\be
\phi_\text{bulk}(r,\varphi_+) \simeq  \frac{S_0}{\Omega}  \Delta\varphi(r)
\ee
where $\Delta\varphi(r)=\varphi_+(r) - \varphi_-(r) \neq 0$. In contrast, the boundary condition \eqref{BC-ellipse} imposes $\phi=0$ along the ellipse. Consequently, diffusion effects cannot be neglected in a thin region near $\varphi_+(r)$, where the boundary-layer contribution $\phi_\text{layer}$ offsets the nonzero value of the bulk solution $\phi_\text{bulk}$.

Near the exit endpoint, $\phi_\text{layer}$ satisfies the homogeneous leading-order equation
\be
\nabla^2 \phi_\text{layer} - \Omega\,\partial_\varphi \phi_\text{layer} \simeq  0.
\ee
The boundary condition for the layer correction is
\be
\label{rotational-drift-large-Pe-layer-BC}
\phi_\text{layer}(\varphi_+) \simeq - \phi_\text{bulk}(\varphi_+).
\ee
Let $\hat{\mbf{n}}$ denote the outward unit normal to the ellipse, and define
\be
u_n = \bu \cdot \hat{\mbf{n}} = \Omega r\,\hat{\bs{\varphi}} \cdot \hat{\mbf{n}},
\ee
where
\be
\hat{\bs{\varphi}} \cdot \hat{\mbf{n}} = \frac{(a^2 - b^2) \sin\varphi \cos\varphi}{\sqrt{b^4 \cos^2\varphi + a^4 \sin^2\varphi }}.
\ee
At the exit endpoint, $u_n>0$, so the rigid rotational drift crosses the boundary in the outward direction. Also, let $d$ denote the inward normal distance from the waveguide boundary, with $d\ge 0$ inside the guide. In the boundary layer, the diffusion operator is dominated by its normal contribution, so that $\nabla^2\simeq \partial_d^2$. The advective term becomes $\bu \cdot \nabla \simeq - u_n\,\partial_d$, since the remaining contributions are subleading in this region. The leading-order layer equation is therefore
\be
\partial_d^2 \phi_\text{layer} + u_n\,\partial_d \phi_\text{layer} \simeq 0.
\ee
The solution that decays into the interior and satisfies the boundary condition \eqref{rotational-drift-large-Pe-layer-BC} is
\be
\phi_\text{layer} \simeq -\frac{S_0}{\Omega} \Delta\varphi(r) e^{-u_n d}.
\ee
The inward normal distance $d$ and the small angular displacement $\varphi_+(r)-\varphi\ge 0$ along a circular drift line are related by
\be
d \simeq r (\hat{\bs{\varphi}} \cdot \hat{\mbf{n}}) [\varphi_+(r)-\varphi],
\ee
so that
\be
\label{rotational-drift-large-Pe-combined}
\phi_\text{layer}(r,\varphi) \simeq - \frac{S_0}{\Omega}
\Delta\varphi(r) e^{ - \Omega r^2( \hat{\bs{\varphi}} \cdot \hat{\mbf{n}} )^2 \left[\varphi_+(r)-\varphi\right]
}.
\ee

Combining Eqs.~\eqref{rotational-drift-large-Pe-bulk} and \eqref{rotational-drift-large-Pe-combined}, we obtain the composite solution in the outer region, near the exit endpoints:
\begin{align}
\label{omega-drift-large-Pe-outer}
\phi_\text{outer}(r,\varphi) \simeq
& \frac{S_0}{\Omega} \Big\{ \varphi - \varphi_-(r)
\nonumber\\[5pt]
& - \Delta\varphi(r) e^{ - \Omega r^2( \hat{\bs{\varphi}} \cdot \hat{\mbf{n}} )^2 \left[\varphi_+(r)-\varphi\right] } \Big\}.
\end{align}
The exponential factor ensures that the composite solution satisfies the Dirichlet condition at the exit endpoint $\varphi=\varphi_+$. Its contribution becomes negligible in the interior, where the composite solution reduces to the bulk solution. The characteristic angular width of the accumulation layer near the exit endpoint is
\be
\delta \varphi_\text{layer} \sim \frac{1}{\Omega r^2 ( \hat{\bs{\varphi}} \cdot \hat{\mbf{n}} )^2},
\ee
which decreases as the waveguide boundary is approached. The corresponding thickness measured normal to the elliptical waveguide boundary and near it is
\be\label{rotational-drift-large-Pe-thickness}
\delta(\varphi)
\sim \frac{1}{|u_n|}
= \frac{1}{\Omega R(\varphi)\, |\hat{\bs{\varphi}} \cdot \hat{\mbf{n}}|}.
\ee
This expression provides a local estimate of the normal boundary-layer thickness and is valid away from the points where $\hat{\bs{\varphi}} \cdot \hat{\mbf{n}}=0$ (the four axis points of the ellipse), where the first-order expansion underlying the estimate breaks down. Away from these points, the normal thickness of the accumulation layer is estimated as
\be
\delta \sim \frac{1}{\Omega a} = \frac{a}{\text{Pe}_\Omega}.
\ee
It is worth noting, however, that the present approach does not by itself determine the orientation of the layer.

In the core region, where $r<b$, the circular drift lines are closed and lie entirely inside the ellipse, so they do not intersect the waveguide boundary. The advection contribution to the core potential cannot be retained at leading order as in \eqref{rotational-drift-large-Pe-bulk-eq}, since this would produce a term linear in $\varphi$, which is incompatible with the single-valuedness of $\phi$. Periodicity therefore requires $\partial_\varphi \phi_\text{core}=0$ at leading order, so advection does not contribute at this order. The leading-order equation consequently reduces to the radial Poisson equation
\be
\frac{1}{r} \frac{d}{dr} \left( r \frac{d\phi_\text{core}}{dr} \right) \simeq -S_0.
\ee
The solution regular at the origin is
\be
\phi_\text{core}(r) \simeq \frac{S_0}{4} \left( C-r^2 \right).
\ee
The constant $C$ is determined by matching to the outer solution \eqref{omega-drift-large-Pe-outer} at $r=b$. Since the outer solution is $\mathcal{O}(S_0/\Omega)=\mathcal{O}(\text{Pe}_\Omega^{-1})$ there, continuity of $\phi$ requires $\phi_\text{core}(b)=\mathcal{O}(\text{Pe}_\Omega^{-1})$. At leading order, this gives $C=b^2$, and hence
\be\label{rotational-drift-large-Pe-core}
\phi_\text{core}(r) = \frac{S_0}{4} \left( b^2 - r^2 \right) + \mathcal{O}(\text{Pe}_\Omega^{-1}).
\ee
Thus, as $\text{Pe}_\Omega\to\infty$, the potential approaches the axisymmetric core solution \eqref{rotational-drift-large-Pe-core} inside the inscribed circle $r<b$. In the outer region $r>b$, the potential is suppressed to $\mathcal{O}(\text{Pe}_\Omega^{-1})$, except within the narrow exit boundary layers \eqref{rotational-drift-large-Pe-thickness}. This behavior is consistent with the field profiles shown in Fig.~\ref{fig:ellipse-ADV}.

The sharp potential gradients near the exit boundary give rise to an enhanced transverse electric field and the corresponding transverse magnetic field. Under rotational advection, the bulk potential accumulates along circular drift lines over an angular interval of order unity, resulting in a characteristic field gradient
\be
|\ep_\text{bulk}| \sim \frac{S_0}{\Omega a} = \frac{S_0 a}{\text{Pe}_\Omega}.
\ee
Near the exit points, however, the potential must adjust to satisfy the Dirichlet condition across a thin boundary layer. The potential variation across this layer is of order $S_0/\Omega $, while its characteristic thickness is $\delta\sim a/\text{Pe}_\Omega$. The resulting field therefore scales as
\be
|\ep_\text{layer}| \sim S_0 a.
\ee
Consequently,
\be
\frac{|\ep_\text{layer}|}{|\ep_\text{bulk}|} \sim \Omega a^2 = \text{Pe}_\Omega.
\ee
Hence, strong rotational advection suppresses the bulk potential gradient while simultaneously producing a localized enhancement near the exit boundary. The enhancement scales with the rotational P\'{e}clet number, reflecting the increasingly sharp spatial adjustment required to satisfy the boundary condition.

\subsection{Combined uniform and rigid rotational drifts}
\label{sec:combined-drift}

In Sec.~\ref{sec:rotational-drift}, we showed that rigid rotational drift does not affect the solution in a circular waveguide with rotationally symmetric Dirichlet boundary conditions. To examine the effects of rotational advection in this geometry, we now consider the simultaneous presence of uniform and rigid rotational drifts. Although the waveguide cross section remains rotationally symmetric, the uniform drift breaks the rotational symmetry of the differential operator, allowing rotational advection to modify the field profile.

We consider the advection--diffusion equation
\be
\nabla^2\phi - \bu_c \cdot\nabla\phi = - S_0,
\ee
with the combined drift field
\be
\bu_c = \bu + \bu_\Omega.
\ee
Here, $\bu = u \hat{\bu}$ denotes the uniform drift along a fixed direction $\hat{\bu}$, generated by the in-plane harmonic axion field $\theta_{xy}^{(u)}$ given by Eq.~\eqref{axion-uniform-drift}. The second contribution, $\mbf{u}_\Omega = \Omega r \hat{\bs{\varphi}}$, is the nonuniform rigid rotational drift generated by $\theta_{xy}^{(\Omega)}$ in Eq.~\eqref{axion-rotational-drift}. The resulting advection--diffusion equation is
\be\label{u-drift-solid-body-rotation}
\nabla^2\phi - (\bu + \Omega r \hat{\bs{\varphi}}) \cdot \nabla\phi = - S_0.
\ee
When $\bu=0$, the solution in the circular waveguide coincides with the pure-diffusion solution \eqref{diffusion-disc-solution}, as discussed in Sec.~\ref{sec:rotational-drift}. A nonzero uniform drift is therefore essential for observing nontrivial effects of rotational advection in a circular domain with rotationally symmetric boundary conditions.

To solve Eq.~\eqref{u-drift-solid-body-rotation}, we exploit the existence of a stagnation point, defined as the point $\br_s$ at which the uniform and rigid rotational drift contributions cancel. Thus, $\br_s$ satisfies $\bu + \Omega r_s \hat{\bs{\varphi}}_s = 0$, yielding
\be
\br_s = \frac{\z \times \bu}{\Omega}.
\ee
We introduce translated coordinates $(\tr,\tphi)$ centered at $\br_s$, related to the original coordinates $(r,\varphi)$ through $\tbr = \br - \br_s$. In this coordinate system, the combined drift field takes the form of rigid rotation about the new origin. The uniform drift is thus absorbed into the coordinate transformation, while the circular waveguide boundary is displaced relative to the new origin.\footnote{The stagnation point lies at a distance $r_s=u/\Omega $ from the guide center. For $u\gg\Omega R$, it lies far outside the guide (e.g., as $\Omega\to0$ at fixed $u$ or $u\to\infty$ at fixed $\Omega$), so the uniform drift dominates throughout the domain. At $\Omega=0$, the stagnation point recedes to infinity, the shifted-coordinate construction is no longer applicable, and the problem reduces to uniform drift.}

In the shifted coordinates, we define
\be
\tilde{\phi}(\tbr) \equiv \phi(\br) = \phi(\tbr+\br_s),
\ee
which satisfies the advection--diffusion equation
\be\label{g-equation}
\tilde{\nabla}^2 \tilde{\phi} - \Omega \frac{\partial \tilde{\phi}}{\partial\tphi} = - S_0,
\ee
where $\tilde{\nabla}$ is the usual differential operator expressed in terms of $(\tr,\tphi)$. The differential operator now contains only rigid rotational advection; the effect of the uniform drift is instead encoded in the boundary condition. The original circular Dirichlet boundary condition becomes
\be
\tilde{\phi}(\tr,\tphi)|_{\tilde{R}(\tphi)}=0,
\ee
where
\be\label{BC-off-axis}
\tilde{R}(\tphi) = \sqrt{R^2-r_s^2 \sin^2(\tphi-\varphi_s)} - r_s \cos(\tphi-\varphi_s).
\ee
The boundary is a circle of radius $R$ centered at $-\br_s$ relative to the origin of the $(\tr,\tphi)$ coordinate system. Thus, the combined-drift problem is equivalent to a rigid-rotational-advection problem with an off-center circular boundary. This formulation is mathematically analogous to the problem of rigid rotational drift in an elliptical domain discussed in Sec.~\ref{sec:rotational-drift}, except that the boundary asymmetry arises here from the displacement of a circular domain rather than from its elliptical shape.

In analogy with Eq.~\eqref{ellipse-phi-unknowns}, the solution to Eq.~\eqref{g-equation} can therefore be expressed as
\begin{align}
\tilde{\phi}(\tr,\tphi) =
& -\frac{S_0 \tr^2}{4} + A_0
+ \sum_{m=1}^\infty \alpha_m \text{Re} \left\{ I_m(k_m \tr) e^{im\tphi}  \right\}
\nonumber\\[5pt]
& + \sum_{m=1}^\infty \beta_m \text{Im} \left\{ I_m(k_m \tr) e^{im\tphi}  \right\},
\end{align}
where $A_0$ is real and the real-valued coefficient pairs $(\alpha_m,\beta_m)$ are defined in terms of the complex coefficients $A_m$ by $A_m=\tfrac{1}{2}(\alpha_m-i\beta_m)$.

\begin{figure*}[t]
  \centering
  \begin{subfigure}{0.18\textwidth}
    \centering
    \includegraphics[width=\linewidth]{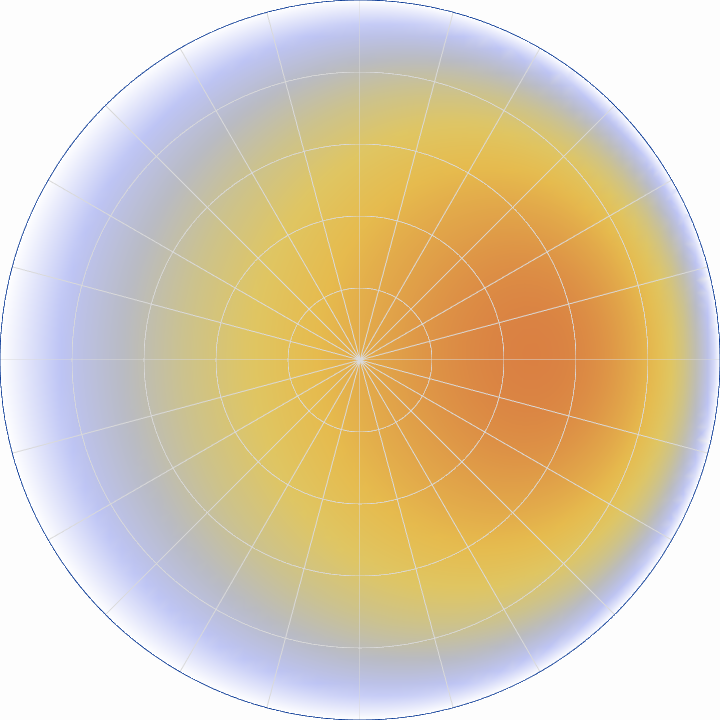}
  \end{subfigure}\hfill
  \begin{subfigure}{0.18\textwidth}
    \centering
    \includegraphics[width=\linewidth]{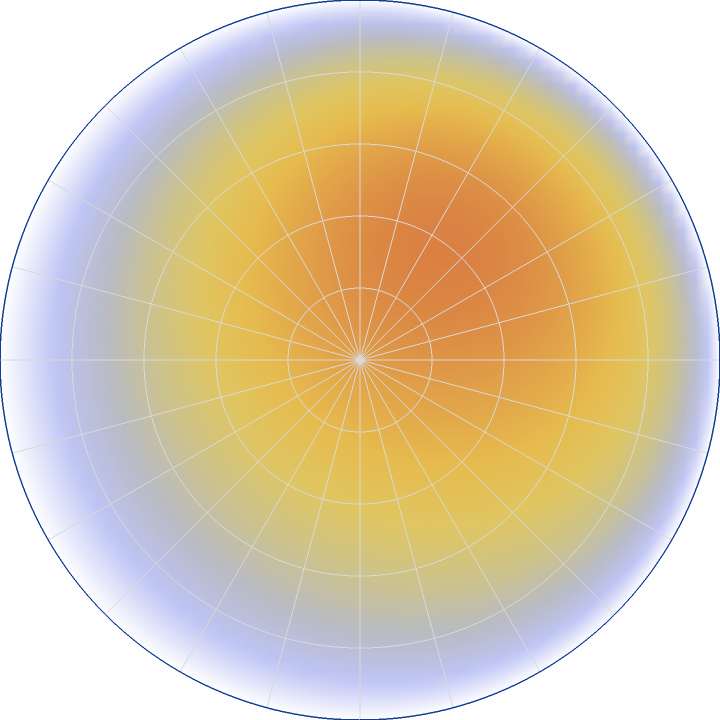}
  \end{subfigure}\hfill
  \begin{subfigure}{0.18\textwidth}
    \centering
    \includegraphics[width=\linewidth]{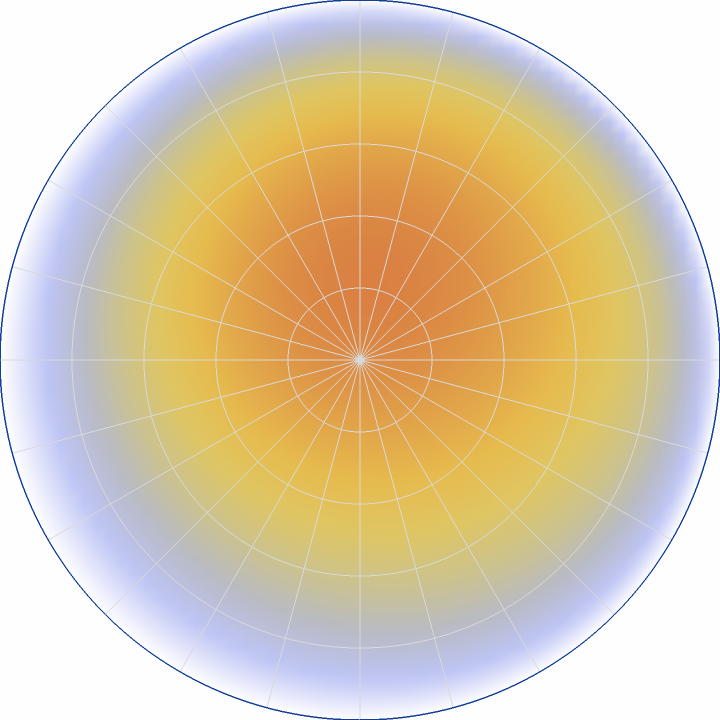}
  \end{subfigure}\hfill
  \begin{subfigure}{0.18\textwidth}
    \centering
    \includegraphics[width=\linewidth]{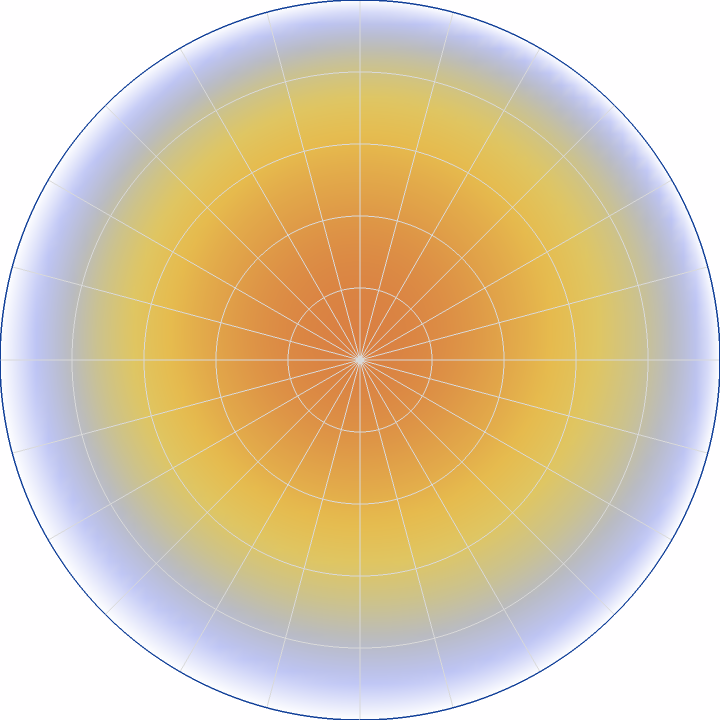}
  \end{subfigure}\hfill
  \begin{subfigure}{0.18\textwidth}
    \centering
    \includegraphics[width=\linewidth]{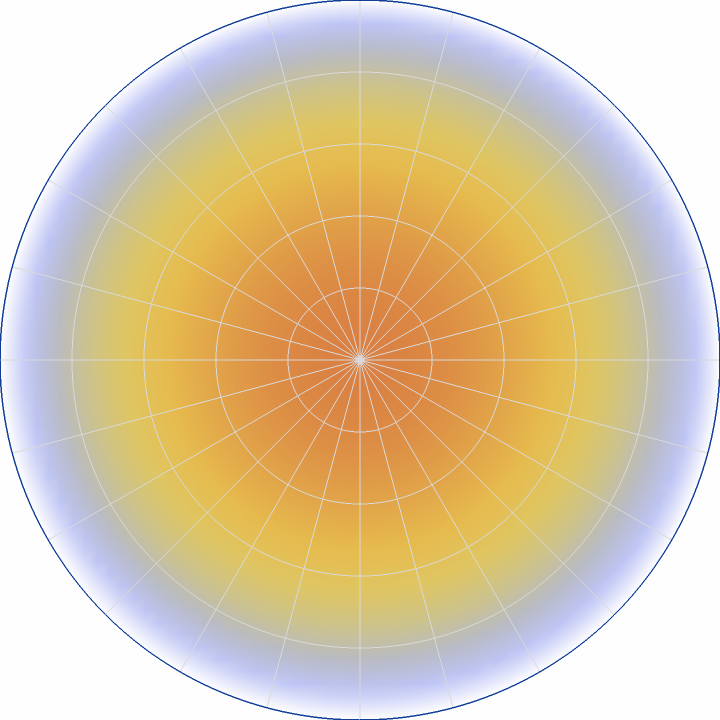}
  \end{subfigure}

\vspace{1em}

  \centering
  \begin{subfigure}{0.18\textwidth}
    \centering
    \includegraphics[width=\linewidth]{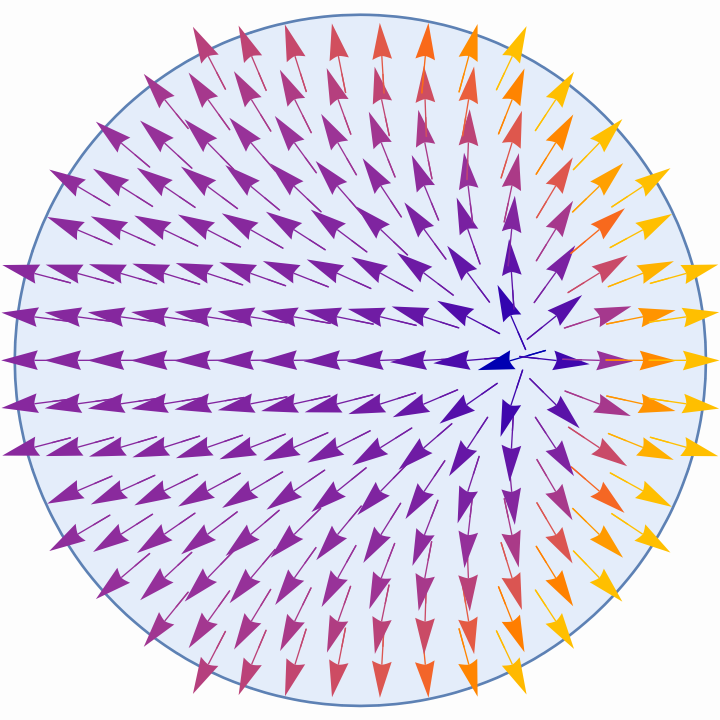}
  \end{subfigure}\hfill
  \begin{subfigure}{0.18\textwidth}
    \centering
    \includegraphics[width=\linewidth]{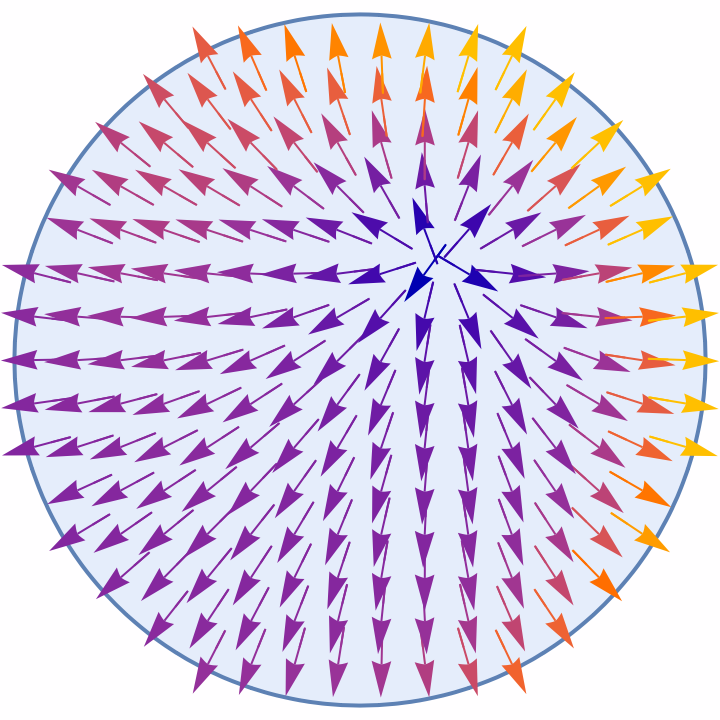}
  \end{subfigure}\hfill
  \begin{subfigure}{0.18\textwidth}
    \centering
    \includegraphics[width=\linewidth]{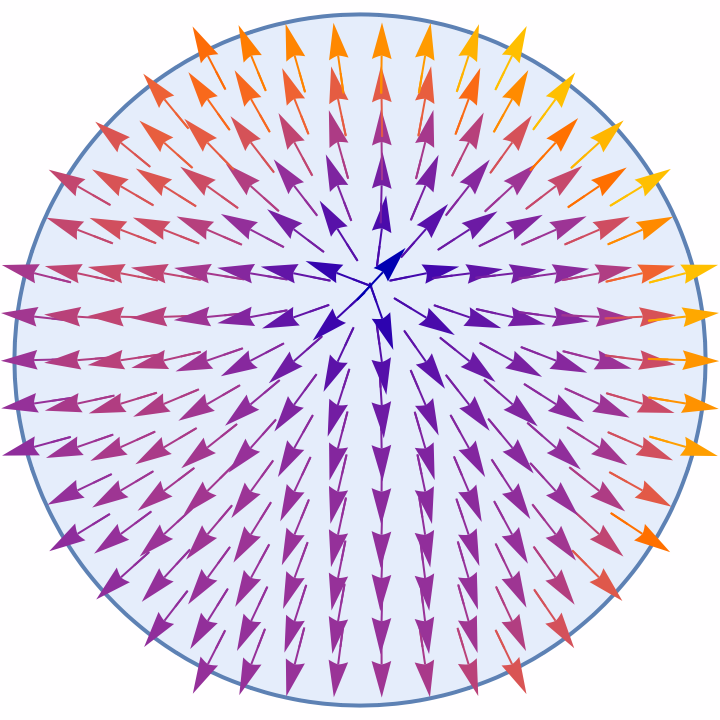}
  \end{subfigure}\hfill
  \begin{subfigure}{0.18\textwidth}
    \centering
    \includegraphics[width=\linewidth]{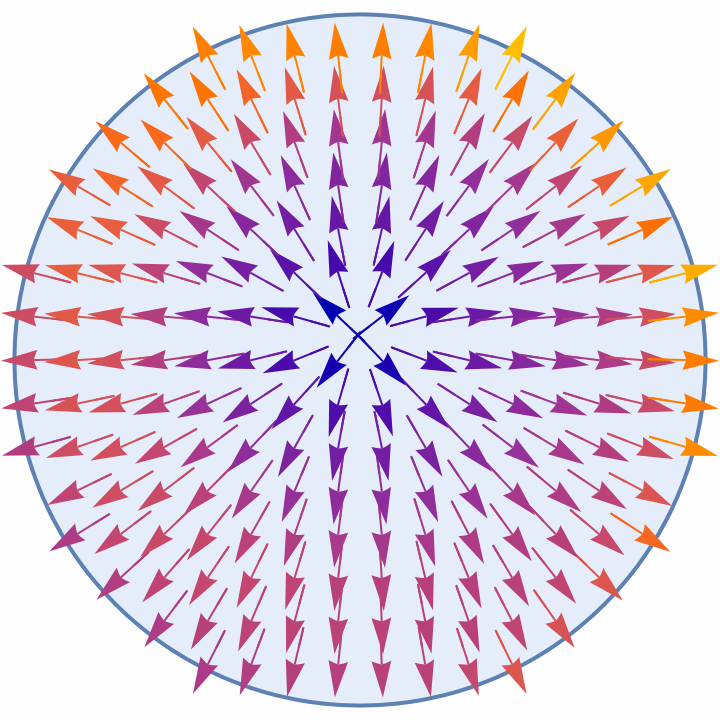}
  \end{subfigure}\hfill
  \begin{subfigure}{0.18\textwidth}
    \centering
    \includegraphics[width=\linewidth]{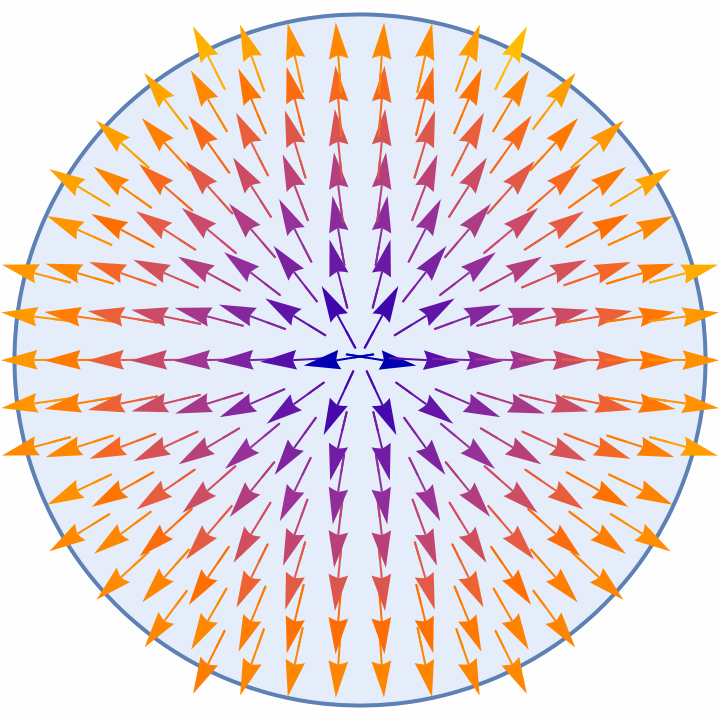}
  \end{subfigure}

\vspace{1em}

\centering
  \begin{subfigure}{0.18\textwidth}
    \centering
    \includegraphics[width=\linewidth]{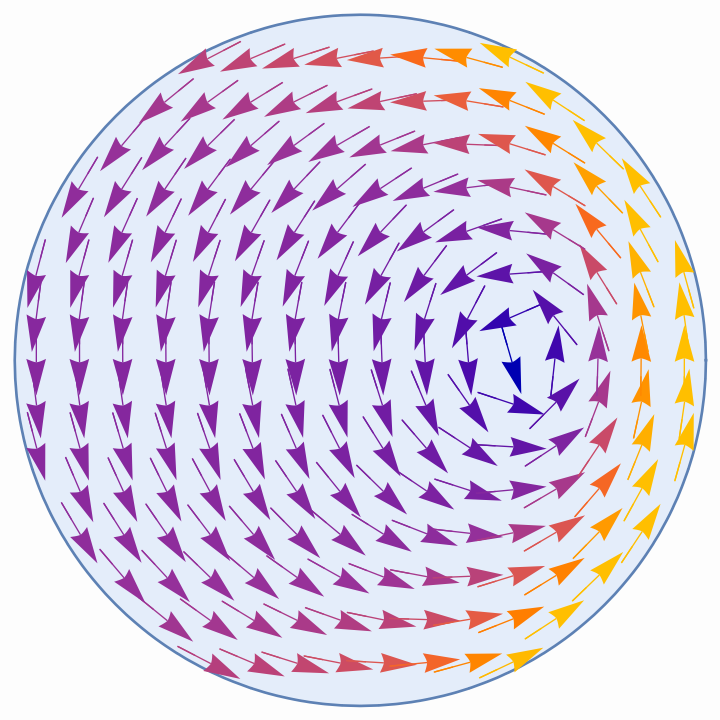}
    \caption{}
    \label{fig:combined-A}
  \end{subfigure}\hfill
  \begin{subfigure}{0.18\textwidth}
    \centering
    \includegraphics[width=\linewidth]{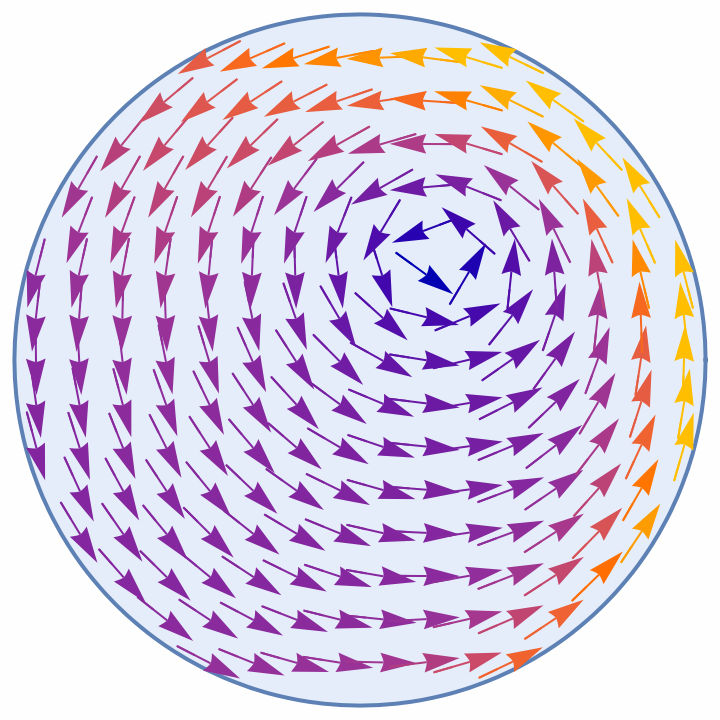}
    \caption{}
    \label{fig:combined-B}
  \end{subfigure}\hfill
  \begin{subfigure}{0.18\textwidth}
    \centering
    \includegraphics[width=\linewidth]{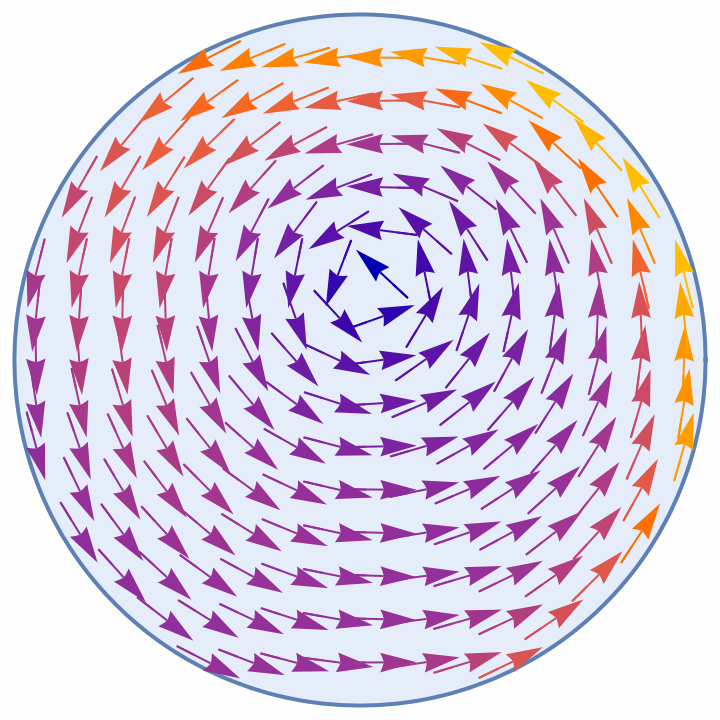}
    \caption{}
    \label{fig:combined-C}
  \end{subfigure}\hfill
  \begin{subfigure}{0.18\textwidth}
    \centering
    \includegraphics[width=\linewidth]{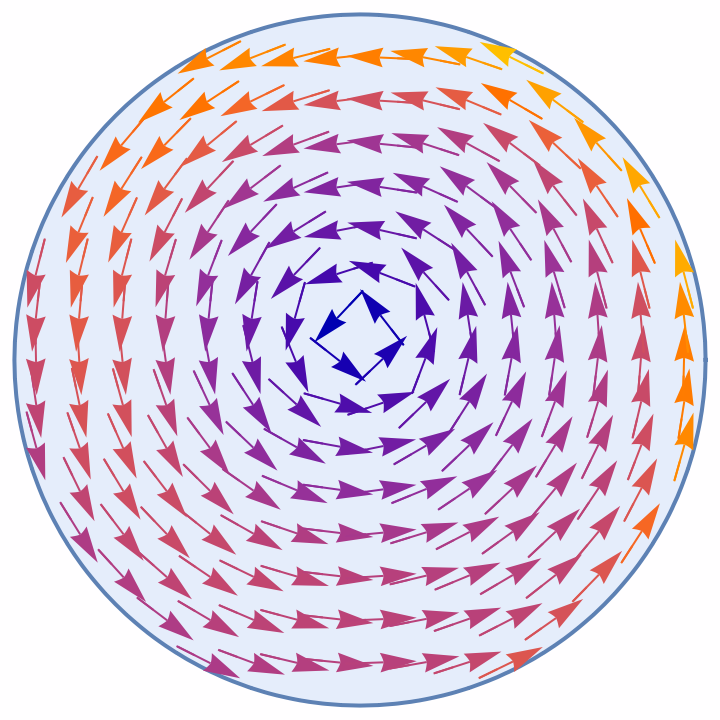}
    \caption{}
    \label{fig:combined-D}
  \end{subfigure}\hfill
  \begin{subfigure}{0.18\textwidth}
    \centering
    \includegraphics[width=\linewidth]{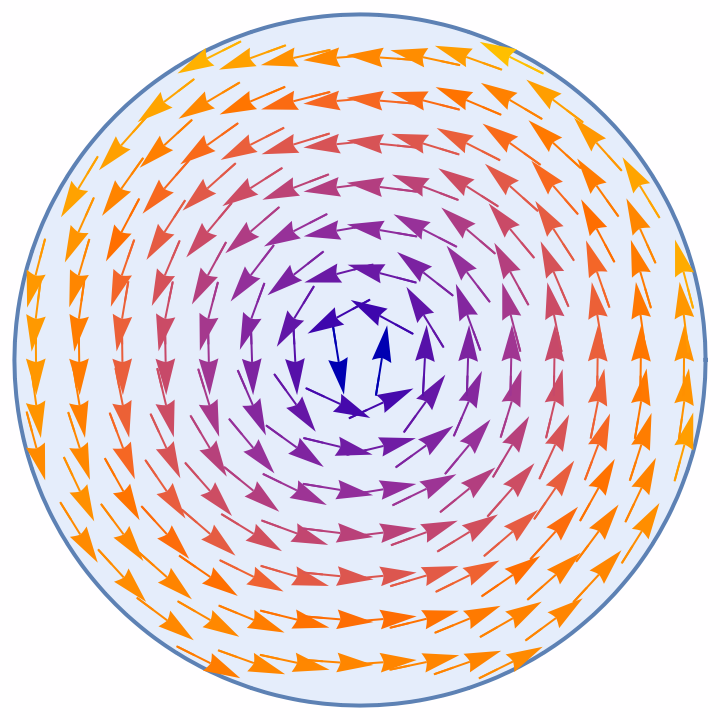}
    \caption{}
    \label{fig:combined-E}
  \end{subfigure}

  \caption{Fields $\phi$ (top), $\ep$ (middle), and $\bp$ (bottom) in the circular guide for the combined uniform drift and rigid rotational drift at $\text{Pe}_\Omega/\text{Pe}_u$ equal to (a) $0.1$, (b) $3$, (c) $6$, (d) $14$, and (e) $20$. Warmer (cooler) colors represent greater (smaller) magnitude.}
  \label{fig:combined-drift}
\end{figure*}

The boundary condition \eqref{BC-off-axis} does not allow the expansion coefficients to be determined directly in closed form, because the boundary radius $\tilde{R}(\tphi)$ depends on the polar angle in the coordinates $(\tr,\tphi)$ used for the expansion. The same difficulty arises when the boundary condition is imposed in the original coordinates: although the boundary is circular in $(r,\varphi)$, expressing the expansion in these coordinates couples the coefficient of the $m$th mode to multiple Fourier components through $\exp\{im\tphi(r,\varphi)\}$. We therefore determine the coefficients $A_0$, $\alpha_m$, and $\beta_m$ numerically, following the collocation procedure described in Sec.~\ref{sec:rotational-drift}. Specifically, we impose the boundary condition on the expansion truncated at angular mode $M$, using $N_\varphi > 2M + 1$ collocation points. The resulting overdetermined linear system is solved by linear least squares. For the plots in Fig.~\ref{fig:combined-drift}, we use $N_\varphi=300$ and $M=35$.

For a characteristic length scale $R$, the relative strengths of uniform and rotational advection with respect to diffusion are characterized by the translational and rotational P\'{e}clet numbers
\be
\text{Pe}_u = uR,
\qquad
\text{Pe}_\Omega = \Omega R^2.
\ee
The ratio
\be
\frac{\text{Pe}_\Omega}{\text{Pe}_u} = \frac{\Omega R}{u}
\ee
instead characterizes the strength of rotational advection relative to uniform advection. Figure \ref{fig:combined-drift} illustrates the behavior of $\phi$, $\ep$ and $\bp$ for different values of this ratio. In the plots, we take the uniform drift along the $x$ axis, $\bu=u\hat{\bx}$ with $u>0$, and consider $\Omega>0$. The stagnation point is then located at $\br_s = (u/\Omega) \hat{\mbf{y}}$. For small values of the ratio, uniform drift dominates the rotational contribution, and the solution approaches the uniform-drift behavior discussed in Sec.~\ref{sec:uniform-drift}. In this regime, $r_s/R \gg 1$. As the rotational drift increases relative to a fixed uniform drift, the stagnation point approaches the center of the circular domain, and the waveguide cross section becomes increasingly centered relative to the effective rotation center. Consequently, the field profile tends toward the rotationally symmetric pure-diffusion solution for $r_s/R \ll 1$, consistent with the behavior discussed in Sec.~\ref{sec:rotational-drift}. At intermediate ratios, the two drift contributions compete: rotational advection changes the orientation of the field profile and progressively distorts it toward a more rotationally symmetric configuration.

\subsubsection{Small-P\'{e}clet-number regime}
\label{sec:combined-drift-small-Pe}

For $\text{Pe}_u\ll1$ and $\text{Pe}_\Omega\ll1$, the combined-drift solution can be expanded systematically in powers of the translational and rotational P\'{e}clet numbers. Introducing the dimensionless radial coordinate $\rho=r/R$ and the normalized potential $\Phi=\phi/(S_0R^2)$, the advection--diffusion equation becomes
\be
\nabla_\rho^2\Phi
-\text{Pe}_u\partial_{\rho_x}\Phi
-\text{Pe}_\Omega\frac{\partial\Phi}{\partial\varphi}
=-1,
\ee
where $\partial_{\rho_x}=\cos\varphi\,\partial_\rho-(\sin\varphi/\rho)\partial_\varphi$. We expand the solution as
\be
\Phi = \Phi^{(0)} + \text{Pe}_u \Phi_u^{(1)} + \text{Pe}_u^2 \Phi_u^{(2)} + \text{Pe}_u \text{Pe}_\Omega \Phi_{u\Omega}^{(2)}
+\cdots.
\ee
The corrections involving only rotational drift vanish at every order because the purely diffusive solution is radial and therefore satisfies the advection--diffusion equation exactly for arbitrary $\Omega$ when $u=0$. The leading terms that depend on the uniform drift are consequently the same as those obtained in Sec.~\ref{sec:uniform-drift-small-Pe} for small P\'{e}clet number, whereas the first contribution involving both drift parameters appears at order $\text{Pe}_u \text{Pe}_\Omega$.

The perturbative hierarchy is
\be
\nabla_\rho^2 \Phi^{(0)} = -1,
\qquad
\nabla_\rho^2 \Phi_u^{(1)} = \partial_{\rho_x} \Phi^{(0)},
\ee
and
\be
\nabla_\rho^2 \Phi_u^{(2)} = \partial_{\rho_x} \Phi_u^{(1)},
\qquad
\nabla_\rho^2 \Phi_{u\Omega}^{(2)} = \frac{\partial\Phi_u^{(1)}}{\partial\varphi},
\ee
with homogeneous Dirichlet boundary conditions at every order. The zeroth-order solution $\Phi^{(0)}$ is the purely diffusive profile \eqref{diffusion-disc-solution}. The first- and second-order uniform-drift corrections, $\Phi_u^{(1)}$ and $\Phi_u^{(2)}$, correspond to the second and third terms in \eqref{u-drift-small-Pe-solution}, respectively. The first mixed correction $\Phi_{u\Omega}^{(2)}$ is determined by the angular derivative of the first-order uniform-drift solution.

Retaining all contributions through second total order in the two P\'{e}clet numbers gives
\begin{align}
\frac{\phi(r,\varphi)}{S_0R^2} =
& \frac{1 - \rho^2}{4} + \text{Pe}_u\, \frac{\rho ( 1 - \rho^2 )}{16} \cos\varphi
\nonumber\\[5pt]
&-\text{Pe}_u^2 \left[ \frac{(1  -\rho^2)^2}{128} - \frac{\rho^2 (1  - \rho^2)}{192} \cos(2\varphi) \right]
\nonumber\\[5pt]
&+\text{Pe}_u \text{Pe}_\Omega \frac{\rho (1 - \rho^2)(2 - \rho^2)}{384} \sin\varphi
\nonumber\\[5pt]
&+\mathcal{O}\left( \text{Pe}_u^3, \text{Pe}_u^2 \text{Pe}_\Omega, \text{Pe}_u \text{Pe}_\Omega^2 \right).
\end{align}
Equivalently, in dimensional variables,
\begin{align}
\phi(r,\varphi) =
& \frac{S_0}{4}(R^2 - r^2) + \frac{S_0 u}{16}\,r(R^2 - r^2) \cos\varphi
\nonumber\\[5pt]
&-S_0u^2
\left[ \frac{(R^2 - r^2)^2}{128} - \frac{r^2 (R^2 - r^2)}{192} \cos(2\varphi) \right]
\nonumber\\[5pt]
& + \frac{S_0 u \Omega}{384} r (R^2 - r^2)(2 R^2 - r^2) \sin\varphi
\nonumber\\[5pt]
& + \mathcal{O}\left( u^3 R^5, u^2 \Omega R^6, u \Omega^2 R^7 \right).
\end{align}
The first three terms recover the purely diffusive solution and the first- and second-order corrections induced by uniform drift. The remaining displayed term is the leading correction arising from the interaction between uniform and rotational advection. This mixed correction has a sine dependence, in contrast to the cosine dependence of the leading uniform-drift correction. It therefore introduces an angular component with a $\pi/2$ phase shift relative to the original dipolar distortion. This behavior is illustrated in Figures \ref{fig:combined-A}-\ref{fig:combined-C}, which show the competition between uniform and rotational advection as $\Omega$ increases by a factor of 60. Since the radial dependence of the mixed term differs from that of the first-order uniform-drift term, the resulting distortion is not, in general, a rigid rotation of the entire field profile.

Thus, although pure rotational drift leaves the solution unchanged in a circular domain, its combination with a weak uniform drift modifies the leading anisotropic correction at order $\text{Pe}_u\text{Pe}_\Omega$. This mixed contribution provides the first perturbative signature of rotational advection in the circular geometry and complements the single-drift expansions derived above.

\subsubsection{Large-P\'{e}clet-number regime}
\label{sec:combined-drift-large-Pe}

\begin{figure}[htb]
	\centering
	\begin{subfigure}{0.5\linewidth}
	  \centering
	  \includegraphics[width=.9\linewidth]{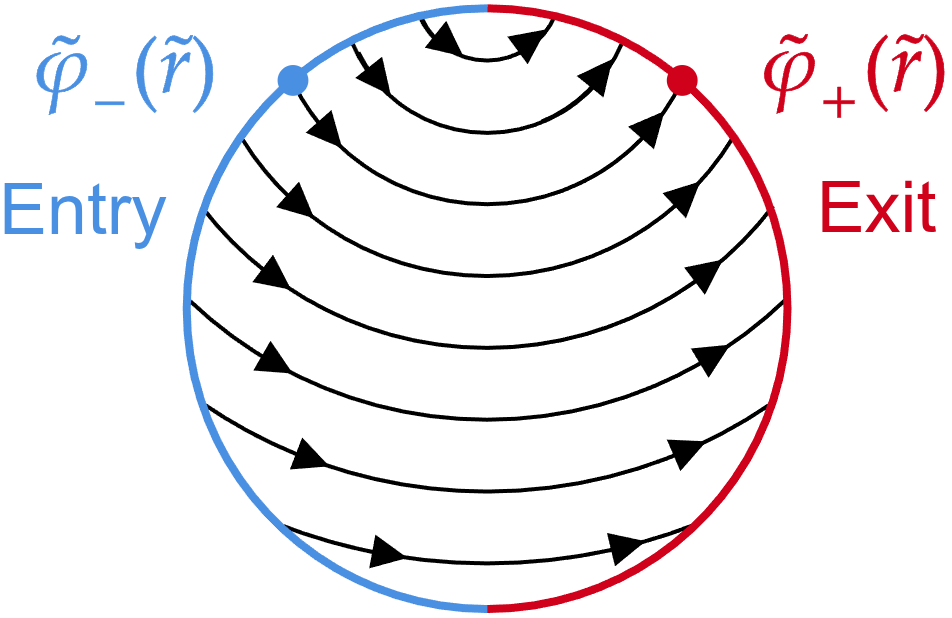}
	  \caption{}
	  \label{fig:stream-combined-A}
	\end{subfigure}\hfill
	\begin{subfigure}{0.5\linewidth}
	  \centering
	  \includegraphics[width=.9\linewidth]{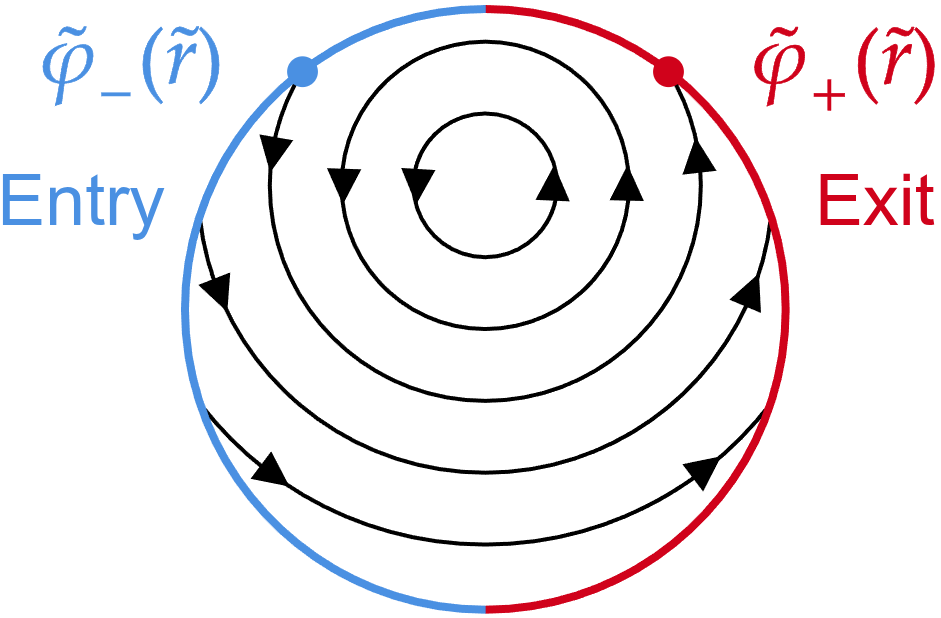}
	  \caption{}
	  \label{fig:stream-combined-B}
	\end{subfigure}\hfill
	\caption{Identification of entry (blue) and exit (red) boundaries of combined uniform and rotational drift lines in the cross section of a circular guide for (a) $r_s > R$ and (b) $r_s < R$.}
	\label{fig:stream-combined}
\end{figure}

We now consider the regime of large translational and rotational P\'{e}clet numbers, $\text{Pe}_u\gg1$ and $\text{Pe}_\Omega\gg1$. In this regime, the relative strengths of the two drift contributions determine the geometry of the drift lines and, consequently, the structure of the potential. The relevant geometric parameter is the distance $r_s=u/\Omega$ between the stagnation point and the center of the circular waveguide, which satisfies
\be
\frac{r_s}{R} = \frac{\text{Pe}_u}{\text{Pe}_\Omega}.
\ee
We first consider the limit in which both P\'{e}clet numbers become large while $r_s/R$ remains fixed. The cases $r_s>R$ and $r_s<R$ exhibit distinct behavior, depending on whether the stagnation point lies outside or inside the waveguide. Figure \ref{fig:stream-combined} illustrates the drift-line geometry in these two cases.

As discussed in Sec.~\ref{sec:combined-drift}, the combined drift field can be written as $\bu_c=\Omega\z\times\tbr$ in a coordinate system centered at the stagnation point $\br_s$, where the drift lines are circles of radius $\tr = |\br - \br_s|$. The displaced circular boundary $\tilde{R}(\tphi)$ is described by Eq.~\eqref{BC-off-axis}. We distinguish two regions: (i) an outer region, where circular drift lines have radii between $R-r_s$ and $R+r_s$ and intersect the boundary; and (ii) a core region, where circular drift lines satisfy $\tr<R-r_s$, are closed, and lie entirely inside the waveguide cross section. A core region exists only when $r_s < R$. In the outer region, let $\tphi_-(\tr)$ and $\tphi_+(\tr)$ denote the entry and exit endpoints, respectively, of the portion of each circular drift line that lies inside the waveguide.

For drift lines that intersect the boundary, the asymptotic decomposition of the outer potential and the advective bulk solution follow the procedure developed in Sec.~\ref{sec:rotational-drift-large-Pe}. We write
\be
\tilde{\phi}_\text{outer} = \tilde{\phi}_\text{bulk} + \tilde{\phi}_\text{layer}.
\ee
Neglecting diffusion in the bulk, Eq.~\eqref{g-equation} reduces to
\be
\partial_{\tphi}\tilde{\phi}_\text{bulk} \simeq \frac{S_0}{\Omega}.
\ee
Imposing the Dirichlet condition at the entry endpoint, $\tilde{\phi}_\text{bulk}(\tphi_-)=0$, gives
\be\label{combined-drift-large-Pe-bulk}
\tilde{\phi}_\text{bulk}(\tr,\tphi) \simeq \frac{S_0}{\Omega} \left[ \tphi - \tphi_-(\tr) \right].
\ee
At the exit endpoint, the bulk solution takes the value
\be
\tilde{\phi}_\text{bulk}(\tr,\tphi_+) \simeq \frac{S_0}{\Omega} \Delta\tphi(\tr),
\ee
where $\Delta\tphi(\tr)=\tphi_+(\tr)-\tphi_-(\tr)$ is the angular extent of the drift line inside the domain. Since this value is generally nonzero, diffusion remains important in a thin accumulation layer near the exit boundary, where the layer correction compensates for the bulk solution.

On the circular waveguide boundary, the rotational drift is tangent to the wall, so the outward normal component of the combined drift is $\bu_c \cdot \hat{\br} = u \cos\varphi$. For convenience, we also introduce the inward normal coordinate $s=R-r$. Thus, near the exit boundary, where $\cos\varphi>0$, the leading-order equation for the layer correction is the same as in Sec.~\ref{sec:uniform-drift-large-Pe}:
\be
\partial_s^2 \phi_\text{layer} + u \cos\varphi\, \partial_s \phi_\text{layer} \simeq 0.
\ee
The resulting layer solution is
\be\label{combined-drift-large-Pe-layer}
\phi_\text{layer}(r,\varphi) \simeq - \frac{S_0}{\Omega} \Delta\tphi(\tr) e^{-u(R-r)\cos\varphi}.
\ee
Combining Eqs.~\eqref{combined-drift-large-Pe-bulk} and \eqref{combined-drift-large-Pe-layer} gives the composite outer solution for $\cos\varphi > 0$:
\begin{align}\label{combined-drift-large-Pe-outer}
\phi_\text{outer}(r,\varphi) \simeq
& \frac{S_0}{\Omega} \left[ \tphi(r,\varphi) - \tphi_-(\tr) \right]
\nonumber\\[5pt]
& - \frac{S_0}{\Omega} \Delta\tphi(\tr) e^{-u(R-r) \cos\varphi}.
\end{align}
At the exit boundary, the two terms cancel to leading order, whereas away from the layer the bulk solution is recovered. The normal thickness of the accumulation layer is therefore
\be
\delta(\varphi) \sim \frac{1}{u \cos\varphi} = \frac{R}{\text{Pe}_u \cos\varphi}.
\ee
As in the uniform-drift problem, this estimate breaks down near $\cos\varphi=0$, where the combined drift is tangent to the waveguide wall. The expression is valid away from these points in the fixed-$r_s/R$ limit, in which both P\'{e}clet numbers become large while their ratio remains finite and nonzero.

When $r_s>R$, the stagnation point lies outside the waveguide, and no closed circular drift line is entirely contained within the domain. All drift lines that intersect the waveguide enter and leave through the boundary. The leading solution is therefore given by Eq.~\eqref{combined-drift-large-Pe-bulk} throughout the interior, with the exit correction \eqref{combined-drift-large-Pe-layer} required to satisfy the Dirichlet condition for $\cos\varphi>0$. There is no closed-drift-line core in this regime.

The limit $r_s\gg R$ corresponds to $\text{Pe}_u\gg\text{Pe}_\Omega$. In this case, the rotational contribution to the drift is small compared with the uniform contribution throughout the waveguide, and the circular drift lines become approximately straight over the domain. The asymptotic solution consequently reduces to the uniform-drift result derived in Sec.~\ref{sec:uniform-drift-large-Pe}, including its bulk profile and exit boundary layer.

When $r_s<R$, the closed-drift-line core requires a separate leading-order solution. As in Sec.~\ref{sec:rotational-drift-large-Pe}, periodicity along the circular drift lines prevents advection from dominating diffusion. The leading-order core potential is therefore independent of $\tphi$ and satisfies the radial Poisson equation
\be
\frac{1}{\tr} \frac{d}{d\tr} \left( \tr \frac{d\tilde{\phi}_\text{core}}{d\tr} \right) \simeq - S_0.
\ee
The solution regular at the origin is
\be
\tilde{\phi}_\text{core}(\tr) \simeq \frac{S_0}{4}(C-\tr^2).
\ee
The constant $C$ is determined by matching to the outer solution \eqref{combined-drift-large-Pe-outer} at $\tr=R-r_s$. Since $\tilde{\phi}_\text{outer} \sim \mathcal{O}(\text{Pe}_\Omega^{-1})$ for a fixed angular extent of the drift line, continuity at the interface requires $\tilde{\phi}_\text{core}(R-r_s)=\mathcal{O}(\text{Pe}_\Omega^{-1})$. At leading order, this gives $C=(R-r_s)^2$, and hence
\be\label{combined-drift-large-Pe-core}
\tilde{\phi}_\text{core}(\tr) = \frac{S_0}{4}\left[ (R - r_s)^2 - \tr^2 \right] + \mathcal{O}(\text{Pe}_\Omega^{-1}).
\ee
The leading-order core profile is therefore parabolic and centered at the stagnation point, with its maximum at $\tr=0$:
\be
\phi_{\max} = \frac{S_0}{4}(R - r_s)^2 + \mathcal{O}(\text{Pe}_\Omega^{-1}).
\ee
The displacement of the stagnation point thus shifts the center of the leading-order potential profile and reduces its maximum relative to the pure-diffusion solution \eqref{diffusion-disc-solution}, as illustrated in Fig.~\ref{fig:combined-D}.

Finally, the limit $r_s \ll R$, or $\text{Pe}_\Omega \gg \text{Pe}_u$, corresponds to strong rotational drift at fixed uniform drift. The field profiles in this regime are shown in Fig.~\ref{fig:combined-E}. This limit requires a separate asymptotic expansion from the fixed-$r_s/R$ regime considered above. In the present limit, the stagnation point approaches the center of the circular waveguide, $r_s = u/\Omega \to 0$, and the leading-order potential approaches the pure-diffusion solution \eqref{diffusion-disc-solution},
\be
\phi_\text{d}(r)=\frac{S_0}{4}(R^2-r^2).
\ee
To determine the leading correction, we write $\phi = \phi_\text{d} + \psi$, retaining terms linear in $u$. Substituting into Eq.~\eqref{u-drift-solid-body-rotation}, with $\bu = u\hat{\bx}$, gives
\be
\nabla^2 \psi - \Omega\partial_\varphi \psi = - \frac{S_0ur}{2} \cos\varphi,
\qquad
\psi(R,\varphi) = 0.
\ee
We seek an asymptotic expansion of the form $\psi = \psi_\text{bulk} + \psi_\text{layer}$. For $\text{Pe}_\Omega\gg1$, the leading-order bulk correction is
\be
\psi_\text{bulk}(r,\varphi) \simeq \frac{S_0ur}{2\Omega} \sin\varphi.
\ee
This bulk solution does not satisfy the Dirichlet condition at $r=R$. A boundary layer is therefore required to cancel its nonzero boundary value. In the limit considered here, the uniform drift is weak compared with the rotational drift, so its normal advective contribution is subleading within the boundary layer. Instead, normal diffusion balances tangential rotational advection. Introducing the inward normal coordinate $s=R-r$, the dominant terms in the equation for the layer correction are
\be
\label{combined-drift-large-Pe-r<<R-layer-eq}
\partial_s^2 \psi_\text{layer} - \Omega \partial_\varphi \psi_\text{layer}
\simeq0.
\ee
The layer correction satisfies the homogeneous leading-order equation because the bulk solution already cancels the source term at leading order. The boundary condition follows from requiring the composite correction to vanish at $r=R$:
\be
\psi_\text{layer}(R,\varphi) \simeq - \psi_\text{bulk}(R,\varphi).
\ee
To solve Eq.~\eqref{combined-drift-large-Pe-r<<R-layer-eq}, we introduce a complex exponential solution proportional to $e^{-\sqrt{i\Omega} s}e^{i\varphi}$. The layer correction is consequently
\begin{align}
\psi_\text{layer}(r,\varphi)
& \simeq - \frac{S_0 u R}{2\Omega} \text{Im} \left\{
e^{-\sqrt{i \Omega}(R - r)} e^{i\varphi} \right\}.
\end{align}
Since $\text{Re}\sqrt{i\Omega}=\sqrt{\Omega/2}$, the corresponding normal thickness scales as
\be
\delta_\Omega \sim \frac{1}{\sqrt{\Omega}} = \frac{R}{\sqrt{\text{Pe}_\Omega}}.
\ee
Combining the pure-diffusion solution with the bulk and boundary-layer corrections, we obtain
\begin{align}
\phi(r,\varphi) \simeq
& \frac{S_0}{4}(R^2-r^2) + \frac{S_0ur}{2\Omega} \sin\varphi
\nonumber\\[5pt]
& - \frac{S_0uR}{2\Omega} \text{Im} \left\{ e^{-\sqrt{i\Omega}(R-r)}e^{i\varphi} \right\}.
\end{align}
At the waveguide boundary, the bulk and layer corrections cancel to leading order, thereby restoring the Dirichlet condition. This approximation is valid when $u\ll\sqrt{\Omega}$, or, equivalently, when
\be
\text{Pe}_u \ll \sqrt{\text{Pe}_\Omega}.
\ee
Thus, when $r_s<R$, the leading-order solution contains a displaced closed-drift-line core, as shown in Fig.~\ref{fig:combined-D}. In contrast, for fixed uniform drift and sufficiently strong rotation, $r_s\ll R$, the bulk potential approaches the rotationally symmetric pure-diffusion profile, as shown in Fig.~\ref{fig:combined-E}.

\subsection{Concluding Remarks}
\label{sec:conclusion}

The analysis presented in this work establishes a connection between TEM propagation in hollow-conductor waveguides filled with axion medium and the mathematical structure of steady advection--diffusion problems. The central result is that the spatial variations of the axion background generate an effective drift field that governs the transverse electrostatic potential and, consequently, the electromagnetic field profiles. This correspondence provides a framework in which the effects of the axion background can be understood in terms of the geometry of an associated transport field.

An important feature of this formulation is that the effective drift field is solenoidal but need not be irrotational. This distinction determines the mathematical structure of the governing equation for the electrostatic potential and the possibilities for its analytical reduction. While irrotational drift fields allow a transformation that eliminates the first-order derivative, subject to an additional condition for obtaining a constant-coefficient modified Helmholtz equation, rotational drift generally forbids such a reduction. Thus, the spatial structure of the axion background determines not only the strength but also the mathematical character of the effective transport process.

The solutions further reveal that the geometry of the drift lines, together with the waveguide boundary, plays a central role in determining the transverse field structure. In particular, the distinction between lines that intersect the boundary and closed drift lines provides a useful way of understanding the different asymptotic regimes. In regions crossed by open drift lines, strong advection produces pronounced field accumulation near the exit boundaries, where thin layers develop to accommodate the Dirichlet boundary condition imposed by the waveguide conducting walls. Conversely, closed drift lines constrain the leading-order potential through periodicity, allowing the formation of an approximately axisymmetric interior profile. These features are particularly evident in the cases of rigid rotational drift and combined uniform and rotational drift, where the location of the stagnation point determines the existence and extent of the closed-line region.

The P\'{e}clet number provides a natural measure of the competition between drift and diffusion, but its role is not restricted to characterizing the relative magnitude of the two terms in the governing equation. The asymptotic analyses show that increasing advection can qualitatively reorganize the transverse solution, producing distinct bulk and boundary-layer structures. In the combined-drift problem, the simultaneous presence of translational and rotational advection introduces an additional geometrical dependence, governed by the relative magnitudes of the corresponding P\'{e}clet numbers. This illustrates how different drift components can cooperate or compete in shaping the electromagnetic fields, even when the waveguide geometry itself remains unchanged.

The external magnetic field also plays a fundamental role in the existence of the nontrivial TEM configurations considered here. Under the homogeneous Dirichlet boundary conditions imposed by the perfectly conducting walls, the source-free advection--diffusion problem admits only the trivial solution. The external-field-induced source is therefore essential for sustaining a nonvanishing transverse electrostatic potential within the present formulation. This result clarifies the distinction between the role of the axion-induced drift, which redistributes the transverse field, and that of the source term, which sustains a nontrivial field configuration.

Finally, the present treatment relies on an idealized axion-electrodynamics description and perfect-conductor boundary conditions. Assessing possible material implementations requires a more detailed treatment of the constitutive response, including the compatibility conditions \eqref{conditions-for-TEM} for TEM propagation and the axion field \eqref{axion}, as well as dispersive and dissipative effects. Materials exhibiting effective axion-electrodynamics responses, including Weyl semimetals, may provide candidates for such investigations, although their suitability must be assessed within a material-specific framework. Extensions to more general cross-sectional geometries and spatially varying source configurations also constitute natural directions for further work.

\subsection{Acknowledgements}

The author thanks Professor Edilberto Oliveira Silva (UFMA) for pointing out an issue with the spatiotemporal dependence of the proposed axion field in an earlier version of the manuscript.

The author used OpenAI's ChatGPT (GPT-6) and DeepSeek's DeepSeek-V4.1-Flash-1017 as assistive tools throughout the development of this work. Their use extended substantially beyond language editing and manuscript preparation, encompassing assistance with the identification and development of the theoretical advection--diffusion framework, the establishment of connections between physical concepts, the formulation of mathematical approaches to the analysis of different Péclet-number regimes, the derivation of results, and the physical interpretation and analysis of the findings. The tools were used under the author's direction through an iterative process of scientific inquiry, in which their proposed developments were critically examined and verified by the author. The author takes full responsibility for the originality, mathematical and physical correctness, and overall scientific content of the manuscript. The author also acknowledges the use of Wolfram Mathematica for generating the plots and Mathcha (\url{https://www.mathcha.io/}) for preparing the schematic drawings.

\bibliography{references}

\end{document}